\documentclass[12pt,preprint]{aastex}

\usepackage{CJK}

\usepackage{epsfig}

\shorttitle{[Fe~{\sc iii}] lines in NGC\,2392}
\shortauthors{Zhang et al.}

\begin{document}

\title{[Fe~{\sc iii}] emission lines in the planetary nebula NGC\,2392}

\begin{CJK*}{UTF8}{gbsn}
\author{Y. Zhang (张泳)$^{1}$, X. Fang (方玄)$^2$, W. Chau$^{1}$, C.-H. Hsia (夏志浩)$^{1}$, \\ X.-W. Liu (刘晓为)$^{2,3}$, 
 S. Kwok (郭新)$^{1}$, and N. Koning$^4$ }

\altaffiltext{1}{Department of Physics, The University of Hong Kong, Hong Kong;
zhangy96@hku.hk}
\altaffiltext{2}{Department of Astronomy, Peking University, Beijing 100871,
China}
\altaffiltext{3}{Kavli Institute for Astronomy and Astrophysics, Peking University, Beijing 100871, China}
\altaffiltext{4}{Department of Physics and Astronomy, University of Calgary, Calgary, Canada T2N 1N4}

\begin{abstract}
NGC\,2392 is a young double-shell planetary nebula (PN). Its intrinsic structure and shaping mechanism are still not fully understood. In this paper we present new spectroscopic observations of NGC\,2392. The slits were placed at two different locations to obtain the spectra of the inner and outer regions. Several [Fe~{\sc iii}] lines are clearly detected in the inner region. We infer that NGC\,2392 might have an intrinsic structure similar to the bipolar nebula Mz\,3, which also exhibits a number of [Fe~{\sc iii}] lines arising from the central regions. In this scenario, the inner and outer regions of NGC\,2392 correspond to the inner lobes and the outer outflows of Mz\,3, respectively. We construct a three-dimensional morpho-kinematic model to examine our hypothesis.  We also compare the physical conditions and chemical composition of the inner and outer regions, and discuss the implications on the formation of this type of PN.
\end{abstract}

\keywords{ISM: abundances --- planetary nebula: individual (\object{NGC\,2392})
               }
\maketitle
\end{CJK*}

\section{Introduction}

\begin{figure*}
\begin{center}
\epsfig{file=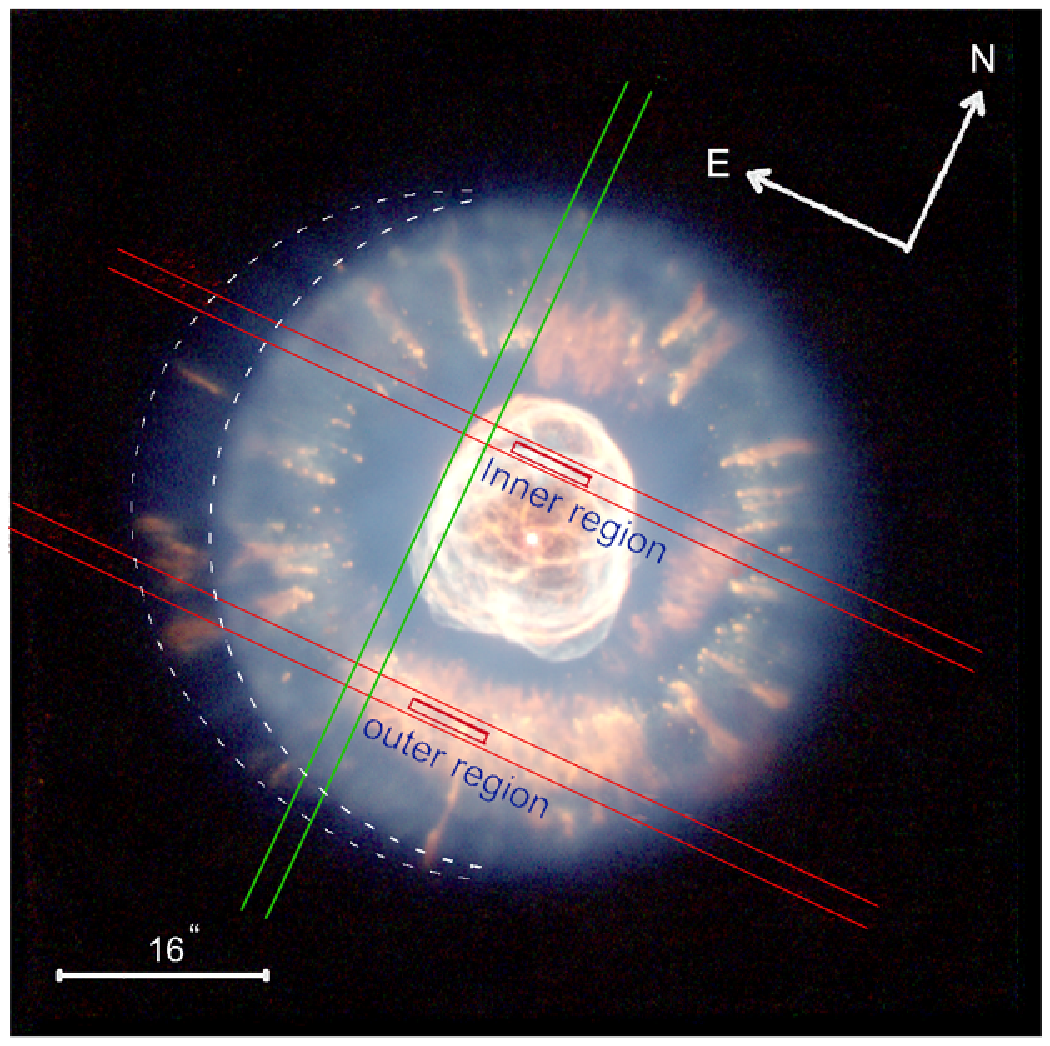, height=15cm}
\end{center}
\caption{Composite-color {\it Hubble Space Telescope} ({\it HST}) images of NGC\,2392. The [O~{\sc iii}] is shown as blue, the H$\alpha$ is shown as green, and the [N~{\sc ii}] is shown as red. The closed boxes represent the positions of the echelle slits. The straight lines give the positions of the YFOSC (red lines) and ESO (green lines) long-slits. The dashed curves outline the extruded arc-shaped structure.}
\label{hst}
\end{figure*}

The Eskimo Nebula, NGC\,2392 (PN G197.8+17.3), is one of the most well studied planetary nebulae (PNe).  It has a distinctive appearance (Figure~\ref{hst}) that is composed of a bright inner shell with a size of $18''\times15''$, and a faint, nearly circular outer shell with a radius of about $23''$. There are several bright rings (or loops) within the inner shell. The faint outer shell contains some bright cometary knots and fuzzy filaments (referred as fuzzes hereafter). Kinematic investigations \citep{reay83,odell85,odell90} have shown that the inner region is expanding with a line-of-sight velocity about four times higher than the outer region. \citet{liller68} found slightly increasing angular velocity with increasing distance from the central star. \citet{gie85}, \citet{balick87}, and \citet{mir90} detected high-velocity bipolar outflows associated with the outer filamentary structure at a position angle of P.A.$=70^\circ$.

The intrinsic morphology of NGC\,2392 is a controversial issue. \citet{weedman68}, \citet{reay83}, and \citet{odell85} modeled this PN with a prolate inner spheroid and a nearly spherical outer shell. \citet{odell90} presented a model where the outer region is an oblate spheroid with the density decreasing away from the central star. \citet{louise81} suggested that the nebula is composed of an inner toroid embedded within
an outer spherical shell. Recently, based on high resolution spectroscopic mapping, \citet{gar11} conclude that the structure of NGC\,2392 is almost the same as NGC\,7009. The nature of the substructures is also highly debated \citep{phi99,odell90,odell02}. It is hard to explain the properties of all the knots and tails in terms of the existing models.

The ionization structure of NGC\,2392 is intriguing. Optical images show that the inner region contains highly-ionized bulges and lowly-ionized loops, and emits lines of ionic species with various ionization states from He~{\sc ii} to [S~{\sc ii}] \citep{odell02}. In the outer region, the high-excitation lines primarily arise from the diffuse disk, while the low-excitation lines originate from the condensations \citep{phi99}. These inner loops and outer condensations appear to be optically thick \citep{dufour11}.  NGC\,2392 also has an unusual thermal structure \citep{liu93a}, which shows outwardly increasing [O~{\sc iii}] and [N~{\sc ii}] temperatures and decreasing \ion{H}{1} Balmer temperatures. \citet{liu93b} and \citet{pei95} found that NGC\,2392 belongs to the group of PNe with extremely large temperature fluctuations. This is probably related to shock heating or the multiple loops of very different ionization structures \citep{liu93b,odell02}. \citet{gue05} detected X-ray emission from the inner region, which suggests the existence of hot plasma at a temperature of $2\times10^6$\,K.

Previous studies have shown that the central star of NGC\,2392 is very bright ($V=10.53$) and has an effective temperature ranging from 32\,000\,K (\ion{H}{1} Zanstra temperature) to 73\,000\,K  (He~{\sc ii} Zanstra temperature) \citep[see the reference in][]{pot08}. It is difficult to understand why the He~{\sc ii} Zanstra temperature is so high.  \citet{dan11} suggested that NGC\,2392 might have a hot, massive white dwarf
companion with a temperature of 250\,000\,K, and might be a Type Ia supernova progenitor. \citet{handler96} found that the central star of NGC\,2392 is variable and may be quasiperiodic with a period of roughly five hours, while the long-term photometry by \citet{she11} suggested that the infrared brightness of the envelope gradually decreases over a period of approximately 2000 days. Compared to normal PNe, the central star of NGC\,2392 has peculiar CNO photospheric abundances \citep{men11}.  

The distance to NGC\,2392 has been derived through various methods, which suggest a value from 0.8\,kpc to 2.1\,kpc \citep[see the reference in][]{odell02,tinkler02}, although \citet{reay83} and \citet{hajian95} suggested a much larger distance ($>4$\,kpc) based on the nebular expansion velocity in the plane of the sky.

The chemical abundances of NGC\,2392 have been determined by several authors \citep[e.g.][]{barker91,henry00,pot08}, based on ultraviolet, optical, and infrared spectra. These studies suggest that this PN was formed in a metal-poor environment and was minimally enriched by nucleosynthesis products. No change in abundance across the nebula is found in any previous study. It is classified as a Type-I PN by \citet{per91} because of the high N/O abundance ratio. The C/O abundance ratio seems to be very uncertain, ranging from 0.38 \citep{barker91} to 1.14  \citep{pot08}.

In this paper, we present new spectra of NGC\,2392, aiming at further investigating the physical conditions, chemical composition, and intrinsic structure of this unusual PN. The observations and data reduction procedures are described in Section~2. In Section~3, we present the detection of iron lines, the plasma diagnostic results, and abundance calculations.  A discussion of the implications of our observations is given in Section~4, followed by the conclusions in Section~5.

\section{Observations and Data Reduction}

\subsection{The YFOSC Observations}

The observations were carried out on December 1--4, 2010 with the Yunnan 2.4\,m telescope, located at Gao Meigu Observatory, China. The telescope was equipped with the Yunnan Faint Object Spectrograph and Camera (YFOSC). A $2000\times4000$ CCD chip was used, leading to a pixel size of 0.28$''$. Long-slit and echelle spectra of two different nebular positions (inner and outer regions) were obtained. A slit width of 1$''$ was used 
throughout the long-slit observations and a $0.58''\times5.37''$ short-slit was used for the echelle observations. The positions of the slits are shown in Figure~\ref{hst}. During the whole observation the slits were oriented in the east-west direction so that the bright loops in the inner region and the south fuzz in the outer region were sampled. The long-slit spectra cover two wavelength ranges: 3300--6000\,{\AA} and 5800--8400\,{\AA} at a resolution of 7\,{\AA} full width at half-maximum (FWHM). The echelle spectra are heavily contaminated by sky emission lines, and the high-order spectra are badly blended together. Therefore, in this study we only use the echelle order covering the wavelength range from 4500--5200\,{\AA}, where the strongest [Fe~{\sc iii}]  lines lie. The echelle spectra have a resolution of 3\,{\AA} FWHM. For each spectrum, three long exposures (1800\,s), and one short exposure (60\,s) were taken.

The spectra were reduced using the IRAF longslit and echelle packages, following the usual steps including subtraction of detector bias, flat fielding, and removal of cosmic rays. He/Ar and Fe/Ar lamps were used for wavelength calibrations and the absolute flux calibration was obtained by observing the standard stars G191-B2B. The one-dimensional spectra were obtained by integrating along the slit. The bright lines falling in the overlapping wavelength region of the red and blue spectra were used to scale the spectra and ensure that they were on the same flux scale. The line fluxes were measured using Gaussian line profile fitting (for weak lines) and direction integration over the observed line profiles (for strong lines), and then were normalized such that H$\beta=100$.

\subsection{The ESO Observations}

Our study is supplemented by the data obtained with the ESO 1.52\,m telescope. The observations were carried out using the long-slit spectrograph Boller \& Chivens (B\&C) on February 1997. The detector was a UV-enhanced Loral $2048\times 2048$ $15\,\mu{\rm m}\times 15\,\mu{\rm m}$
chip.  The B\&C spectrograph has a useful slit length of about 3.5$'$. In order to reduce the CCD read-out noise, the CCD was binned by a factor
of two along the slit direction, resulting in a spatial sampling of 1.63$''$ per\,pixel projected on the sky. A slit width of 2$''$ was used throughout the observations. The slit was oriented in the south-north direction and through the eastern loops in the inner region, as shown in Figure~\ref{hst}. We obtained two long exposures (1800\,s) and one short exposure (180\,s), covering a wavelength range from 4000--5000\,{\AA} at a resolution of 4.5\,{\AA} FWHM. The process of data reduction is the same as described above.

\section{Results}

\subsection{The YFOSC Spectra}

\begin{figure*}
\begin{center}
\epsfig{file=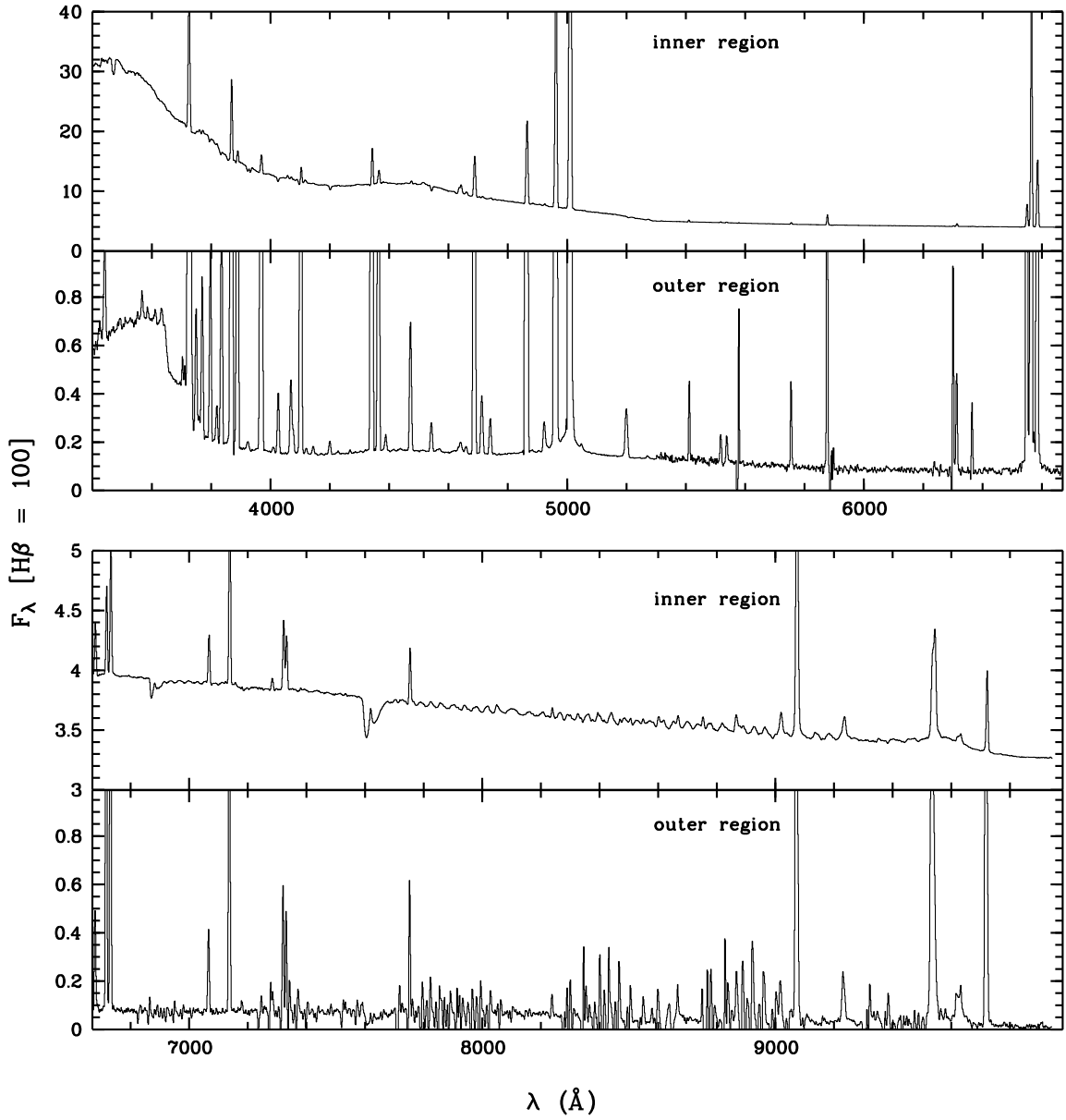,
height=16cm}
\end{center}
\caption{The YFOSC long-slit spectra of NGC\,2392.}
\label{long}
\end{figure*}

\begin{figure*}
\begin{center}
\epsfig{file=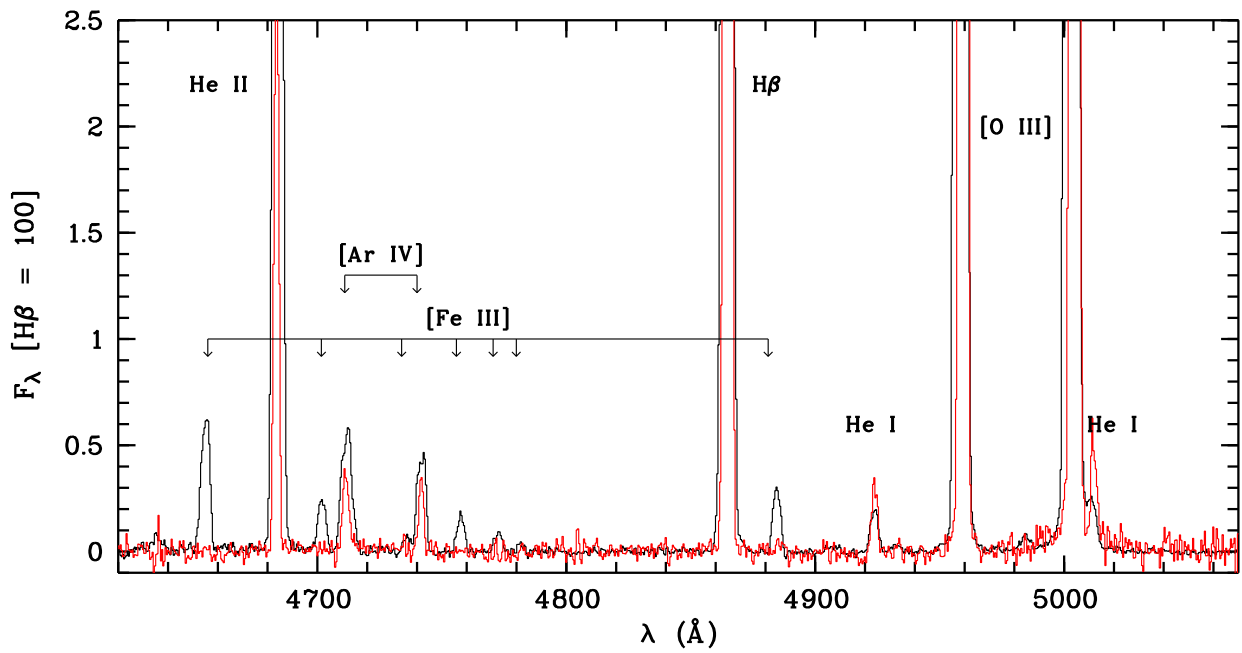,
height=8.5cm}
\end{center}
\caption{The YFOSC echelle spectra of the inner (black) and outer (red) regions of NGC\,2392. Note that [Fe~{\sc iii}] lines are visible only in
the inner region.}
\label{ech}
\end{figure*}

Figures~\ref{long} and \ref{ech} show the long-slit and echelle spectra, respectively. A total of 76 atomic lines (including blended lines) 
belonging to 17 species are detected in the outer region. The detected elements include He, C, N, O, Ne, Ar, S, Cl, and Fe, all of which, except iron, are 1st--3rd row elements in the Periodic Table. Tables~\ref{lline} and \ref{eline} give the detected lines and their intensities
in the long-slit and echelle spectra, respectively. An inspection of Figure~\ref{long} shows that the long-slit spectrum of the inner region
suffers from significant contamination from scattered stellar light, and thus fewer emission lines are detectable compared to that of the outer regions. This effect is particularly remarkable in the blue wavelength region. Except for some [Fe~{\sc iii}] lines (see Section~\ref{feiiiline} for the details), all the lines in the inner region have been detected in the outer region.  

The reddening of NGC\,2392 is low and there is evidence showing that it is inhomogeneous over the nebular surface \citep{zipoy76,dufour11}. From the  H$\gamma$/H$\beta$ and H$\delta$/H$\beta$ flux ratios of the outer region, we obtain a logarithmic extinction at H$\beta$ of $c({\rm H}\beta)=0.27$, in good agreement with the value of 0.22 derived by \citet{liu93b}, but two times higher than that derived by \citet{barker91}. This is probably because the long-slit spectrum primarily samples the dense regions where the extinction by the local dust grains is larger.  However, the  H$\alpha$/H$\beta$ intensity ratio suggests $c({\rm H}\beta)=0$. This might be in part due to the fact that the red and blue spectra, where the H$\alpha$ and H$\beta$ lines respectively lie, were obtained during different nights, and the slit positions for the two wavelength regions are not necessarily the same.  In order to avoid the complexity caused by inhomogeneous extinction, we did not make a reddening correction and simply adopted $c({\rm H}\beta)=0$, as used by \citet{henry00}.  This will introduce an error of up to $20\%$ in the intensities of the lines whose wavelengths are far from that of H$\beta$.

Although \citet{barker91} and \citet{henry00} have presented spectra from several positions of NGC\,2392, their observations are less deep, and only detected relatively strong lines.  The long slit in the outer region is located close to those used for ``Position~5'' of \citet{barker91} and ``Position~A'' of \citet{henry00}, so that we can compare the spectrum of the outer region with those previously obtained in the corresponding positions. We find that the fluxes of strong lines in our spectra are in excellent agreement with those obtained by \citet{barker91} and \citet{henry00}, suggesting a good reliability of the flux measurements.

The line intensities of the long-slit and echelle spectra are not necessarily the same since the ionization structures of NGC\,2392 are spatially varied and the long-slit spectra contain emission from more extended regions (specially the diffuse disk; Figure~\ref{hst}). For example, in the outer region the He~{\sc ii} $\lambda4686$ line of the long-slit spectrum has a flux about four times stronger than that of the echelle spectrum, while the fluxes of the He~{\sc i} $\lambda4921$ line are comparable in the long-slit and echelle spectra. This can be ascribed to the fact that the fuzz has a low-ionization state and He~{\sc ii} lines mainly arise from the diffuse disk.

The echelle spectrum (Table~\ref{eline}) shows that the inner region emits stronger He~{\sc ii} and [Ar~{\sc iv}] lines and weaker He~{\sc i} lines than the outer region, suggesting that the inner loops have a higher excitation than the outer fuzzes. This is consistent with the results by \citet{barker91}. This tendency is less apparent in the long-slit spectra (Table~\ref{lline}) since the spectrum of the inner region bears contamination from the outer region. On the other hand, the long-slit spectra show that, unlike most of the other lines, the [\ion{O}{1}] lines are particularly strong in the outer region. This is consistent with the imaging observations by \citet{phi99} who found that the [\ion{O}{1}]/[O~{\sc iii}] map is enhanced in the south. [\ion{N}{1}] emission is also clearly detected in the outer region, but not in the inner region.
It follows that the south fuzz may contain neutral gas.

Moreover, the [S~{\sc ii}] lines behave differently in the inner and outer regions. The two transitions from 2D$^{\rm o}$ to 4S$^{\rm o}$ in the outer region are about two times stronger than those in the inner regions, whereas the [S~{\sc ii}] 2P$^{\rm o}$--4S$^{\rm o}$ transitions do not show such a trend. This can also be found in the spectra of \citet{barker91} and \citet{henry00}. We infer that the [S~{\sc ii}] lines primarily originate from the south fuzz where the electron temperatures are relatively low and thus the 2D$^{\rm o}$--4S$^{\rm o}$ transitions are easier to excite. The temperature-sensitive line [O~{\sc iii}] $\lambda4363$ does not show such behavior because the [O~{\sc iii}] lines arise mainly from the diffuse nebula where the electron temperatures are relatively homogeneous. \citet{phi99} indeed found that the [S~{\sc ii}] image is more clumpy than the [O~{\sc iii}] image.

\subsection{The ESO Spectrum}

\begin{figure*}
\begin{center}
\epsfig{file=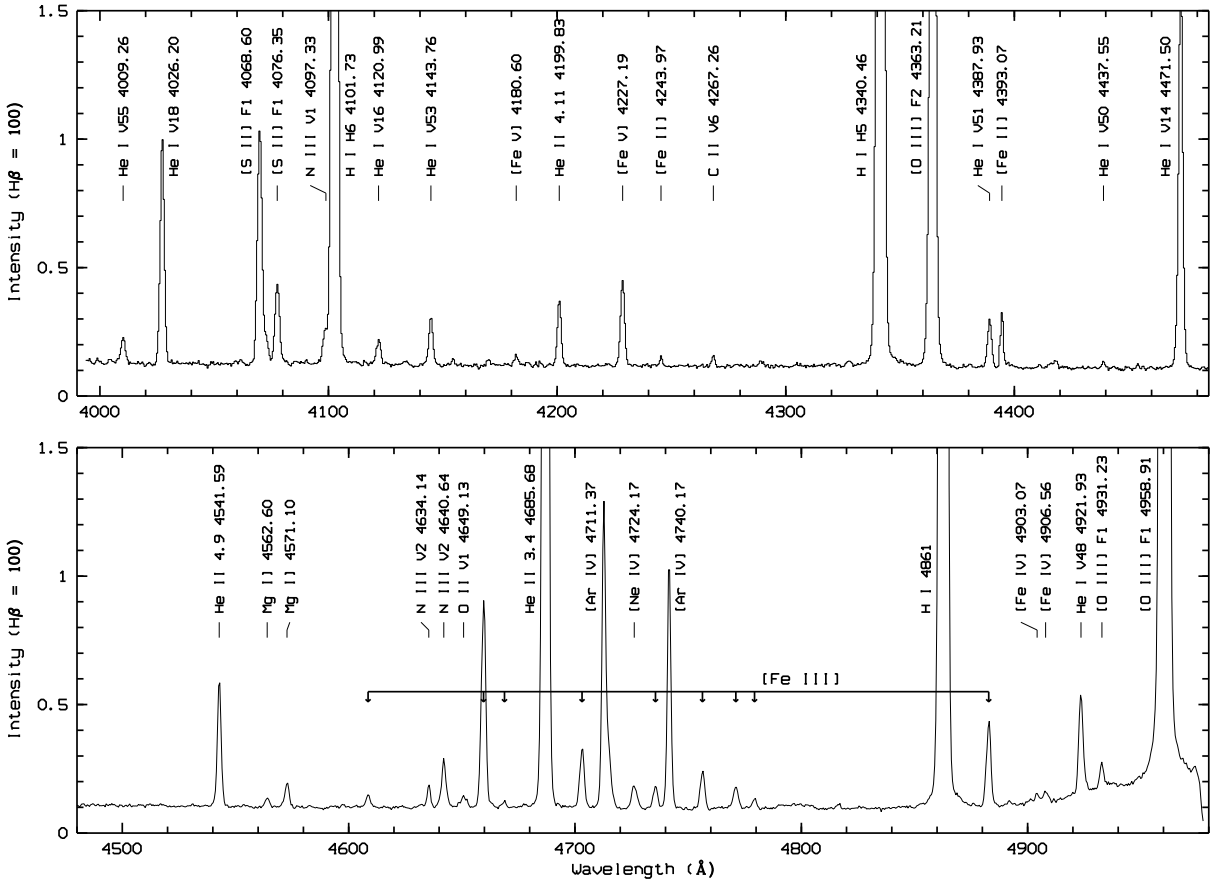,
height=16cm, angle=-90}
\end{center}
\caption{The ESO spectrum of NGC\,2392. }
\label{esospe}
\end{figure*}

The ESO spectrum is dominated by emission from the inner region. Although the wavelength coverage is narrower, the ESO spectrum has a higher signal-to-noise ratio than the YFOSC spectra. Figure~\ref{esospe} shows the ESO spectrum as well as the line identifications.  We detect a total of 44 individual features. Using the H$\beta$, H$\gamma$, and H$\delta$ line fluxes, we obtained an average $c({\rm H}\beta)$ value of 0.08. The de-reddened line fluxes are given in Table~\ref{esolinelist}. The relative fluxes of detected lines only slightly differ from those of the inner regions obtained by YFOSC.

\subsection{The [Fe~{\sc iii}] Lines \label{feiiiline}}

The most striking result of our observations is the detection of ten [Fe~{\sc iii}] lines (Tables~1--3), which were not detected by 
\citet{barker91} and \citet{henry00} because of their lower instrumental sensitivities. All the [Fe~{\sc iii}] transitions are from the lowest excitation levels to the ground state (from $^3$F, $^3$H, $^3$D, to $^5$D), and are the strongest ones within the wavelength range investigated.
 \citet{del09} investigated the iron abundance in a sample of PNe including NGC\,2392, and detected five of the [Fe~{\sc iii}] lines in this PN.
Compared with their spectra \citep[see Figure~1 in][]{del09}, our detections obviously have much higher signal-to-noise. 

Since the intensity ratio of two transitions originating from the same upper level simply depends on their spontaneous transition probabilities
instead of the physical conditions of the nebula, we can use such transitions to test the theoretical calculations of transition probabilities as well as the reliability of our measurements.  We have detected five pairs of [Fe~{\sc iii}] lines arising from the same upper levels.  The most recent calculations of [Fe~{\sc iii}] transition probabilities were presented by \citet{qui96} and \citet{nah96} using different methods. Such calculations bear some uncertainties because some approximations need to be made to obtain the wavefunctions. In Table~\ref{comp} we compare the observed intensity ratios and the theoretical values. They are in good agreement, confirming the reliability of the theoretical calculations and our detections. The flux ratios from the ESO spectrum exhibit better agreement with the theoretical  predictions, implying more reliable flux measurements.

The behavior of these [Fe~{\sc iii}] lines is particularly intriguing. It is clear from the echelle spectra (Figure~\ref{ech}) that these [Fe~{\sc iii}] lines originate overwhelmingly from the highly-ionized inner region, while the echelle spectrum of the outer region does not show any [Fe~{\sc iii}] lines at all. This distinguishes the [Fe~{\sc iii}] lines from the others. While the two strongest [Fe~{\sc iii}] lines at 4658\,{\AA} and 5270\,{\AA} are also detected in the long-slit spectra of the outer region, they are extremely faint.  Compared to the other lines, those of [Fe~{\sc iii}] have the largest intensity contrast between the inner and outer regions (see Table~\ref{lline}).

\begin{figure*}
\begin{center}
\epsfig{file=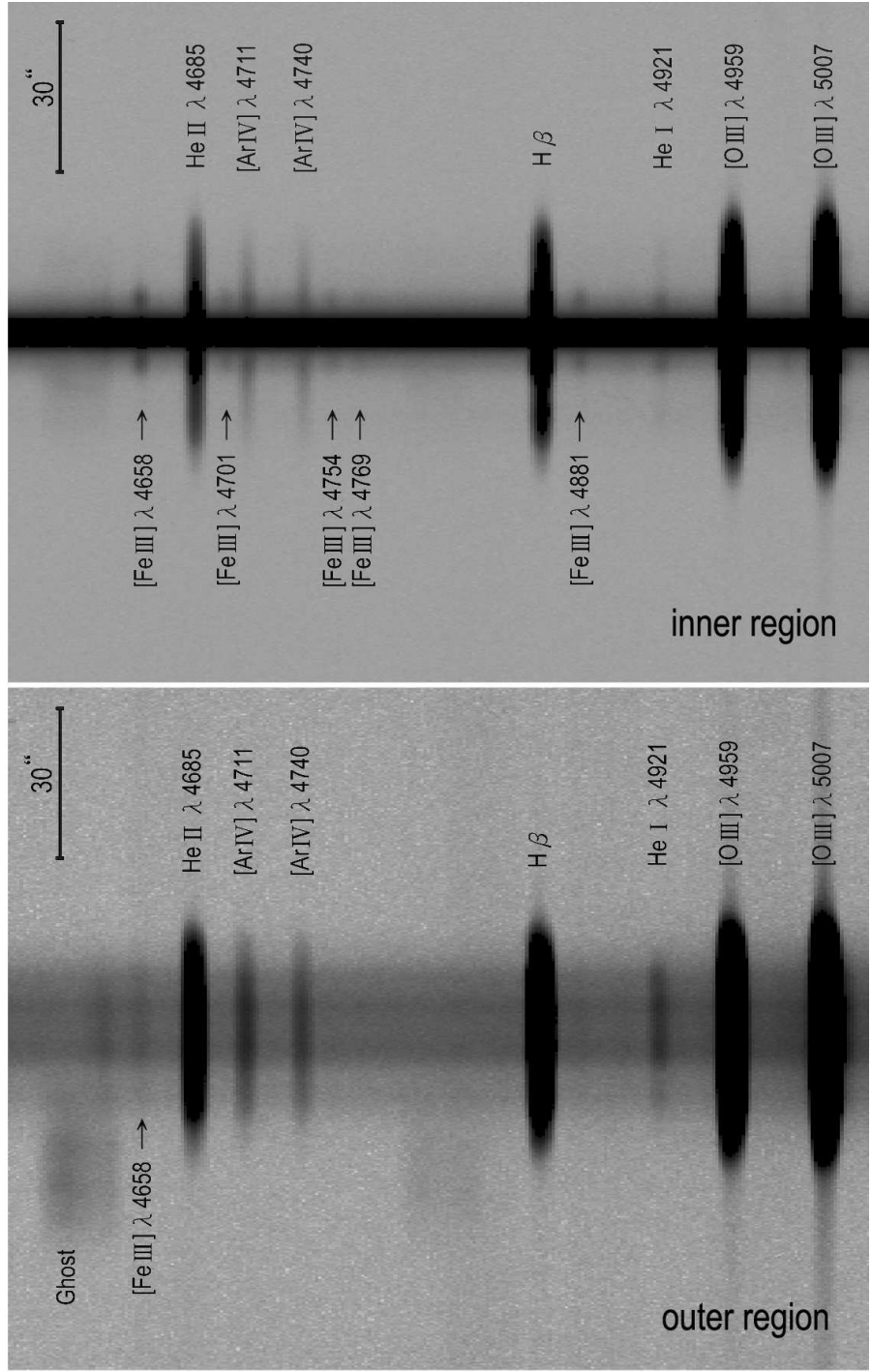,
height=19cm}
\end{center}
\caption{The two-dimensional YFOSC long-slit spectra of NGC\,2392.
 }
\label{2d}
\end{figure*}

The spatial distribution of [Fe~{\sc iii}] lines can also be revealed in the two-dimensional long-slit spectra (Figure~\ref{2d}; also see Figure~\ref{hst} for the slit positions). The inner-region spectrum shows that all the [Fe~{\sc iii}] lines are less spatially extended than 
the others.  In the outer-region spectrum, most of the [Fe~{\sc iii}] lines are invisible. Such a spatial distribution cannot be attributed to the change of excitation states as other low-ionization species do not exhibit such behavior.  Therefore, we conjecture that iron may be greatly 
enhanced in the inner region. 

In our previous work \citep[][ZL02 hereafter]{zhang02}, we detected rich and prominent [Fe~{\sc iii}] emission lines in the bipolar PN Mz\,3 (PN G331.7--01.0). The [Fe~{\sc iii}] lines in Mz\,3 are about one order of magnitude stronger than those in NGC\,2392.  We found that analogous to those in NGC\,2392, the [Fe~{\sc iii}] lines originate exclusively from the central region of Mz\,3. Both NGC\,2392 and Mz\,3 are young PNe and likely have a binary central star.  X-ray emission has been detected in the central regions of both PNe \citep{gue05,kastner03}.  Therefore, we propose that the two PNe might share a common origin (see Section~\ref{dis} for further discussion). 

\subsection{Plasma Diagnostics}

\begin{figure*}
\begin{center}
\epsfig{file=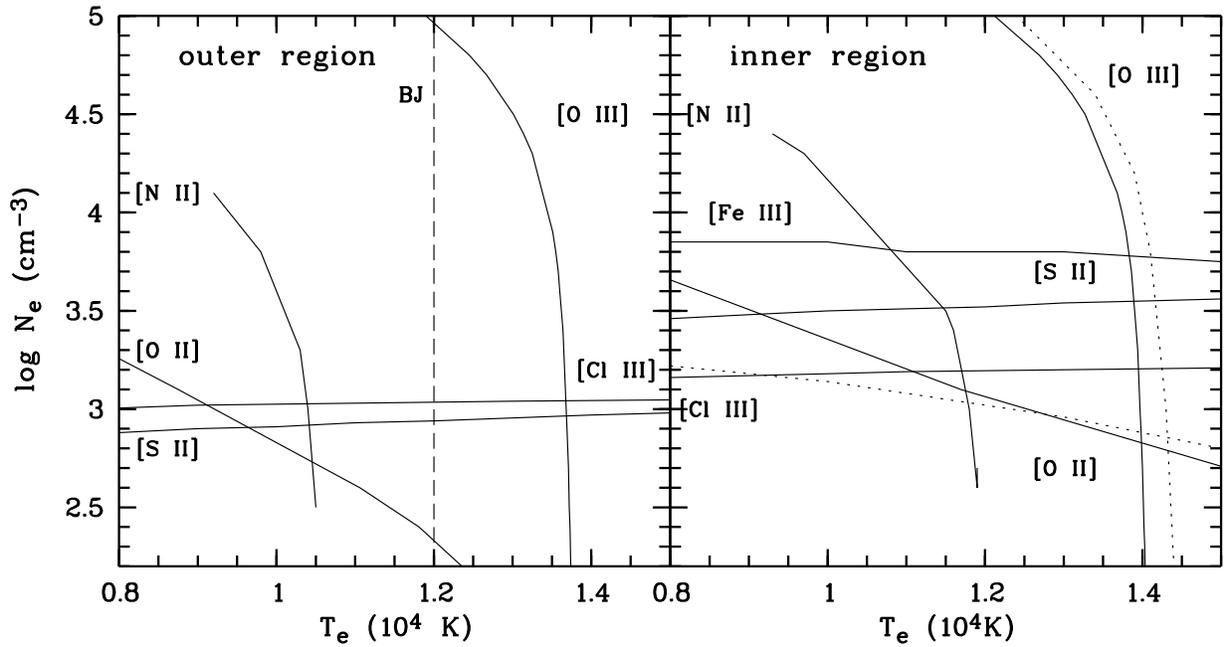,
height=8.5cm}
\end{center}
\caption{Plasma diagnostic diagram. The diagnostics are labeled by the ions. [S~{\sc ii}] represents {[S~{\sc ii}]} $\lambda6731/\lambda6716$. [Cl~{\sc iii}] represents {[Cl~{\sc iii}]} $\lambda5537/\lambda5517$. [O~{\sc ii}] represents [O~{\sc ii}] ($\lambda7320 + \lambda7330)/(\lambda3726+\lambda3729)$. {[N~{\sc ii}]} represents [N~{\sc ii}] $(\lambda6548 + \lambda6584)/\lambda5754$.  [O~{\sc iii}] represents {[O~{\sc iii}]} $(\lambda4959 + \lambda5007)/\lambda4363$. [Fe~{\sc iii}] represents [Fe~{\sc iii}] $\lambda4658/\lambda4701$. The dash line denotes the temperature derived from the Balmer jump. The diagnostics are based on the YFOSC observations except for the dotted lines which are obtained by the [O~{\sc iii}] and [Fe~{\sc iii}]  lines detected in the ESO spectrum. }
\label{dia}
\end{figure*}

Plasma diagnostics are carried out using the same method and atomic data as those described in ZL02. Figure~\ref{dia} shows the plasma diagnostic
diagrams of the inner and outer regions. The resultant electron temperatures and densities are given in Table~\ref{diagnostic}. The plasma diagnostics are mainly based on the collisionally excited lines detected in YFOSC long-slit spectra except for those based on [Fe~{\sc iii}] lines which are taken from the echelle spectra. We also determine the electron temperature in the outer region using the nebular continuum Balmer jump at 3643\,{\AA} ($T_{\rm BJ}$). However, it is impossible to accurately measure the Balmer jump in the inner region due to strong contamination from the scattered stellar light.

As shown in Figure~\ref{dia} and Table~\ref{diagnostic}, in both regions the electron temperatures determined by the [O~{\sc iii}] lines ($T_{\rm OIII}$) are higher than those by the lowly-ionized lines, suggesting that the lowly-ionized clumpy structures are cooler. A comparison between $T_{\rm OIII}$ and $T_{\rm BJ}$ can provide information about the temperature fluctuations inside the nebula \citep{pei67}. We find that the $T_{\rm BJ}$ value of the outer region is only 1700\,K lower than $T_{\rm OIII}$, implying insignificant temperature fluctuations in this region. Our results also show that the inner region has higher electron densities compared to the outer region. 

ZL02 have invoked some [Fe~{\sc iii}] line pairs to derive the electron temperatures and densities. Given their high critical densities, [Fe~{\sc iii}] lines are particularly useful to probe the high density regions. Using [Fe~{\sc iii}] lines, ZL02 found that the central region of Mz\,3 has an abnormally high density ($\sim10^6$\,cm$^{-3}$). In the spectra of NGC\,2392, the temperature-sensitive [Fe~{\sc iii}] lines, which arise from upper levels with high excitation energies, are too weak to be detected. We determine the electron density of the inner region using only one pair of [Fe~{\sc iii}] lines. The results show that the electron density derived from the [Fe~{\sc iii}] lines is about 2--4 times higher than those derived from other lines. Therefore, [Fe~{\sc iii}] lines might originate from slightly denser regions. However, we do not find an extremely dense region equivalent to that of Mz\,3.

In Figure~\ref{dia}, we also plot the diagnostic results using the [O~{\sc iii}] $\lambda4959/\lambda4363$ ratio based on the ESO observations. The curve closely resembles that obtained from the YFOSC observations, suggesting that these inner loops have a similar $T_{\rm OIII}$ value. The two curves obtained from the [O~{\sc iii}] and [Fe~{\sc iii}] line ratios converge to a point where $T_{\rm e}=14600$\,K and $N_{\rm e}=10^{2.8}$\,cm$^{-3}$.

\begin{figure*}
\begin{center}
\epsfig{file=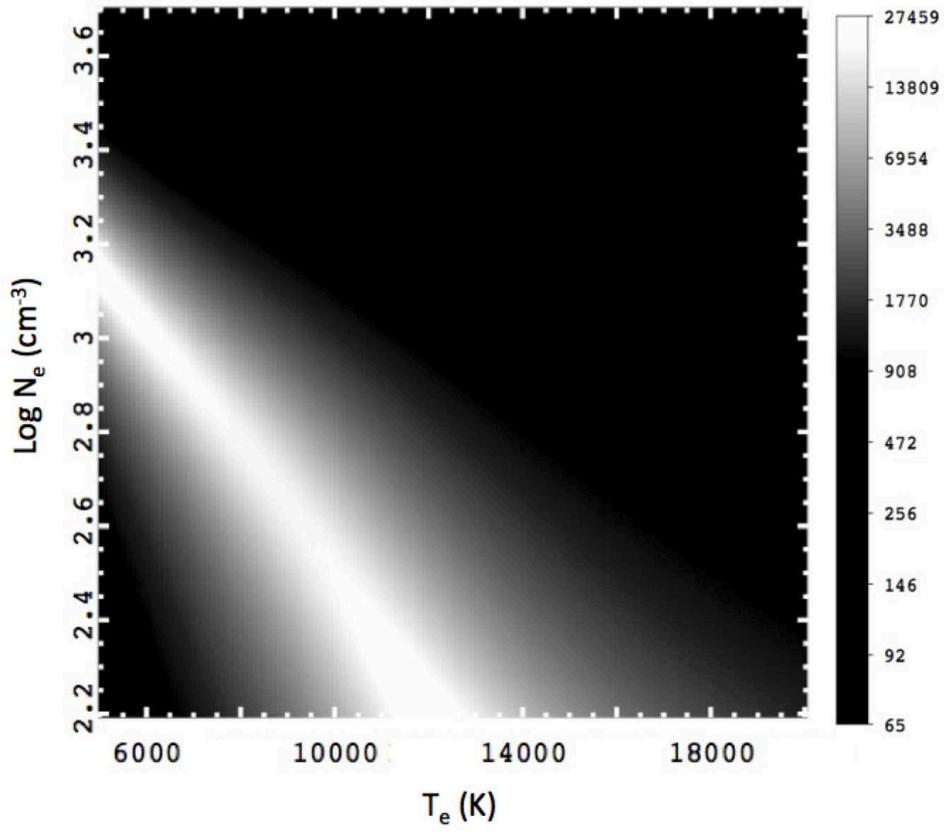,
height=11cm}
\end{center}
\caption{The diagnostic diagram of [Fe~{\sc iii}] lines in the ESO spectrum (see the text). The  $1/\chi^2$ level is indicated by the color bar.
}
\label{kai}
\end{figure*}

Nine [Fe~{\sc iii}] lines are detected with reliable fluxes by the ESO spectrum. Based on a Chi-square analysis, we determine the nebular physical conditions utilizing all the [Fe~{\sc iii}] lines. For this, we first normalize the intensities of all the [Fe~{\sc iii}] lines such that $I_{\rm obs}(\lambda_{4658})=1$. We define $\chi^2(T_{\rm e},N_{\rm e})$ as the sum of the squares of the intensity-weighted deviation between the observed and predicted intensities, $\chi^2(T_{\rm e},N_{\rm e})=\sum_i[I_{(T_{\rm e},N_{\rm e})}(\lambda_{i})-I_{\rm obs}(\lambda_{i})]^2\times I^2_{\rm obs}(\lambda_{i})$, where $I_{(T_{\rm e},N_{\rm e})}$ is the predicted intensity under a given electron temperature and density.  Figure~\ref{kai} displays the distribution of 1/$\chi^2$ in the $T_{\rm e}$---$N_{\rm e}$ space. Although the minimum $\chi^2$ value does not converge to definite $T_{\rm e}$ and $N_{\rm e}$ values because of the lack of $T_{\rm e}$-sensitive [Fe~{\sc iii}] lines, it is clear from this figure that if $T_{\rm e}$ has a normal value (5000--20000\,K) these [Fe~{\sc iii}] lines arise from regions with $N_{\rm e}<10^{3.2}$\,cm$^{-3}$. This confirms the above conclusion that there is no extremely dense region similar to the one found in Mz\,3.

\subsection{Abundances}

\subsubsection{YFOSC Observations}

We calculate the ionic and elemental abundances using the same procedures described in ZL02. For the ionic abundance calculations, we adopt  $T_{\rm e}=10000$\,K and 12000\,K, and $\log N_{\rm e}$(\,cm$^{-3}$)$=3.0$ and 3.5 for the outer and inner regions, respectively. Note that the abundance calculations of He$^+$, He$^{2+}$, and C$^{2+}$ are based on recombination lines and thus are not sensitive to the assumed electron temperatures and densities. Table~\ref{abund} gives the derived ionic abundances.  

The abundances of most of the ions are in agreement in the two regions within one order of magnitude. The most notable difference is that the Fe$^{2+}$ abundance in the inner region is about 15 times higher than that in the outer region. Furthermore, the outer region has an O$^0$/H$^+$ abundance ratio about nine times higher than the inner region. We cannot make direct comparisons with previous results because of the different slit locations. We would like to mention that \citet{pot08} and \citet{del09} respectively obtained Fe$^{2+}$/H$^+=8.0\times10^{-7}$ and $4.2\times10^{-7}$, lying in the middle of the values of the inner and outer regions derived in the present paper.

In Table~\ref{element} we present the elemental abundances. For comparison, we also list the values previously derived by other authors \citep{pot08,henry00,barker91} and the solar photospheric abundances \citep{lodders}. The He abundances are the sum of He$^+$/H$^+$ and  He$^{2+}$/H$^+$. Following ZL02, we use the S$^+$ and S$^{2+}$ abundance ratio to estimate the ionization correction factor (ICF) of carbon.
The Cl abundance of the outer region is the sum of Cl$^+$/H$^+$ and Cl$^{2+}$/H$^+$. In the inner region, Cl$^+$ is not detected and we use the ICF suggested by ZL02 to obtain the Cl abundance. The ICFs for nitrogen, oxygen, neon, and argon are obtained using the expressions suggested
by \citet{kb94}. For iron, \citet{pot08} did not make the ionization correction. Based on photoionization models and an empirical analysis, \citet{rod05} presented two schemes to derive the ICF for iron. According to their methods, we obtain ICF(Fe)$=1.86$ and 2.00 for the outer region, and ICF(Fe)$=2.34$ and 2.57 for the inner region. The average values are used to calculate the total iron abundances.

As shown in Table~\ref{element}, the abundances derived by different authors are diverse. Previous studies show large uncertainties of the C/O ratio \citep{pot08}, ranging from 0.12 to 1.14. Our results suggest a C/O ratio of 0.33.  The nitrogen abundance is much lower than those
derived by other authors. For nitrogen, we only observe one ionic species N$^+$. We note that the N$^+$/H$^+$ abundance ratios in both regions are comparable to that derived by \citet{pot08}. Therefore, the ICF of nitrogen might suffer from some uncertainties.  Considering the differences in the adopted atomic data, electron temperatures, densities, and ICF schemes, it is not surprising that the abundances derived by different authors are not the same. Furthermore, given the complex structures of NGC\,2392, it is almost impossible to obtain accurate ICFs.

Nevertheless, when comparing the relative abundances between the inner and outer regions, the uncertainties introduced by the abundance
calculations will be greatly reduced. We find that except for iron there is no appreciable difference in the elemental abundances between the outer and inner regions. The iron abundance in the inner region is about 15 times higher than that in the outer region. The abundance discrepancy clearly cannot be accounted for by the errors in the measurements and calculations, and implies an enhancement of iron in the inner region.

Although the plasma diagnostics do not suggest very high electron densities, it is arguable that some spatially unresolved knots consisting of  high-density gas may exist in the inner region, and the electron density derived by the [Fe~{\sc iii}] line ratio is the mean value of the whole region. If this was the case, one may argue that the enhancement of [Fe~{\sc iii}] lines is not due to the higher iron abundance in the inner region but the presence of dense knots ($N_{\rm e}\sim10^6$\,cm$^{-3}$) where the lines other than [Fe~{\sc iii}] are substantially collisionally de-excited.  However, in order to achieve pressure balance, the knots are required to have an electron temperature lower than the surrounding gas
by a factor of $\sim10^3$. In such a cold environment, iron cannot stay in gas phase because of its high condensation temperature. Consequently, we can rule out the possibility that the enhancement of [Fe~{\sc iii}] lines is caused by the existence of dense knots in the inner region.

\subsubsection{ESO observations}

Compared to the YFOSC spectrum of the inner region, the ESO observations show a similar electron temperature and slightly lower electron density. For the calculations of ionic abundances, we adopted $T_{\rm e}=12000$\,K and 14600\,K for the lowly- and highly-ionized species respectively, and a constant electron density of $10^{2.8}$\,cm$^{-3}$. Table~\ref{eso_abund} gives the results. The ionic abundances obtained from the ESO observations agree with those derived from the YFOSC within one order of magnitude. We do not attempt to derive the elemental abundances because of the small number of ionic species detected in the limited wavelength coverage. 

Extensive studies of PNe have consistently shown that the abundance of heavy elements derived from collisionally excited lines are systematically lower than those derived from optical recombination lines, and these discrepancies are positively correlated with the $T_\mathrm{e}$([O~{\sc iii}])-$T_\mathrm{e}$(H~{\sc i}~BJ) values \citep[defined as the ``abundance problem'', see][for a review on this subject]{liu06}. Heavy element recombination lines, C~{\sc ii} and O~{\sc ii}, are detected with low intensities in the ESO spectrum, allowing us to investigate the abundance problem in this PN. The strongest O~{\sc ii} line at 4649\,{\AA} is blended with two weaker lines. The peaks of the three lines can be clearly seen (Figure~\ref{esospe}). We decompose the three components by multiple Gaussian fitting and find that 52.4$\%$ of the intensity of the feature is due to the O~{\sc ii} $\lambda4649$. The O$^{2+}$ abundances derived from recombination and collisionally excited lines suggest a low abundance discrepancy factor (ADF) of 1.65. Since the ADF values of PNe typically range from 1.6--3.2 and positively correlate with the temperature fluctuations \citep{liu06}, the low ADF value and large temperature fluctuations of NGC\,2392 make this PN unusual. \citet{liu06} suggests that the C/O abundance ratio derived from the two types of lines are comparable, and thus the C/O ratio derived from C~{\sc ii} and O~{\sc ii} recombination lines is more reliable. Under the assumption that C/O$=$C$^{2+}$/O$^{2+}$, the recombination lines result in a C/O ratio of 0.305.

\section{Discussion \label{dis}}

Previous studies have shown that the abundance of iron in PNe is significantly lower than the solar value \citep{per99,del09}. This has been commonly attributed to the condensation of Fe into dust grains. The depletion factors of iron (Fe$_\sun$/Fe$_{\rm PN}$) are very diverse, ranging from 10--1000, and are independent of nebular properties. On the other hand, the destruction of dust grains might release considerable Fe into the gas phase. Therefore, the study of iron abundance can shed light onto dust processes. According to the abundance calculations of NGC\,2392, we obtain depletion factors of 9 and 166 in the inner and outer regions, respectively. In this section we shall discuss possible interpretations of the different Fe depletion factors in the two regions.

\subsection{Are the Eskimo and the Ant Twin PNe?}

\begin{figure*}
\begin{center}
\epsfig{file=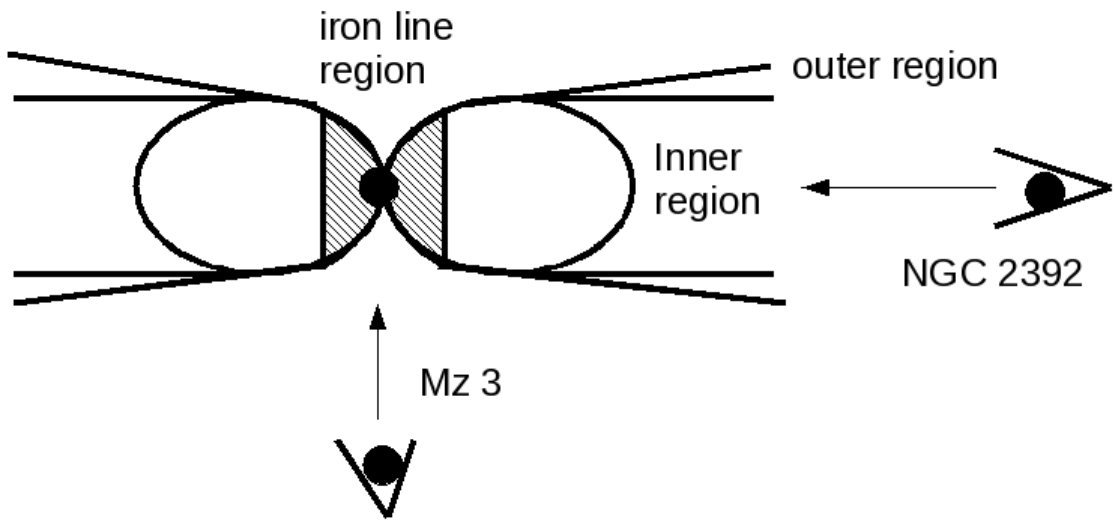,
height=10cm}
\end{center}
\caption{Schematic drawing of a bipolar nebula, showing our claim that NGC\,2392 and Mz\,3 have the same intrinsic structure, but are viewed from different perspectives.}
\label{sch}
\end{figure*}

The behavior of iron abundances in Mz\,3 (ZL02) is similar to that in NGC\,2392. The emission lines of nickel in Mz\,3 also exclusively arise from the central region \citep{zhang06}. It is suggested that the gas in the central region of Mz\,3 was expelled from a giant companion, and thus shows a different abundance pattern compared to the outer outflows. Here we suggest that NGC\,2392 likely shares a common origin with Mz\,3, and
the inner region has been recently launched by a giant companion. Mz\,3 has a bipolar structure, which is often interpreted in terms of the interaction of a binary system \citep[e.g.][]{soker98}. It follows that NGC\,2392 might share the same intrinsic structure with Mz\,3.
Figure~\ref{sch} illustrates our idea that both NGC\,2392 and Mz\,3 are bipolar PNe, but viewed in pole-on and edge-on directions respectively.

\begin{figure*}
\begin{center}
\epsfig{file=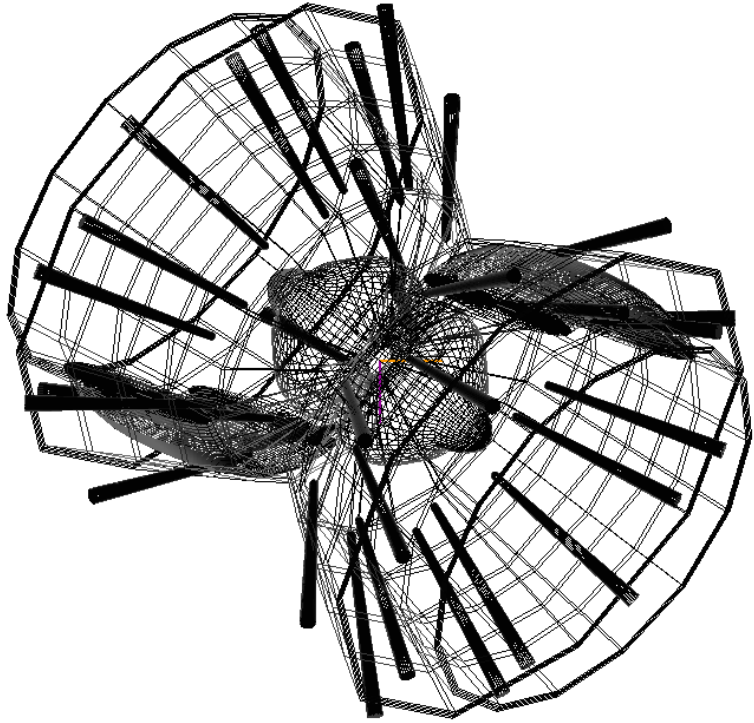,
height=10cm}
\end{center}
\caption{The three-dimensional mesh geometry of our model. }
\label{mesh}
\end{figure*}

\begin{figure}
 \begin{center}
 \begin{tabular}{cc}
\resizebox{70mm}{!}{\includegraphics{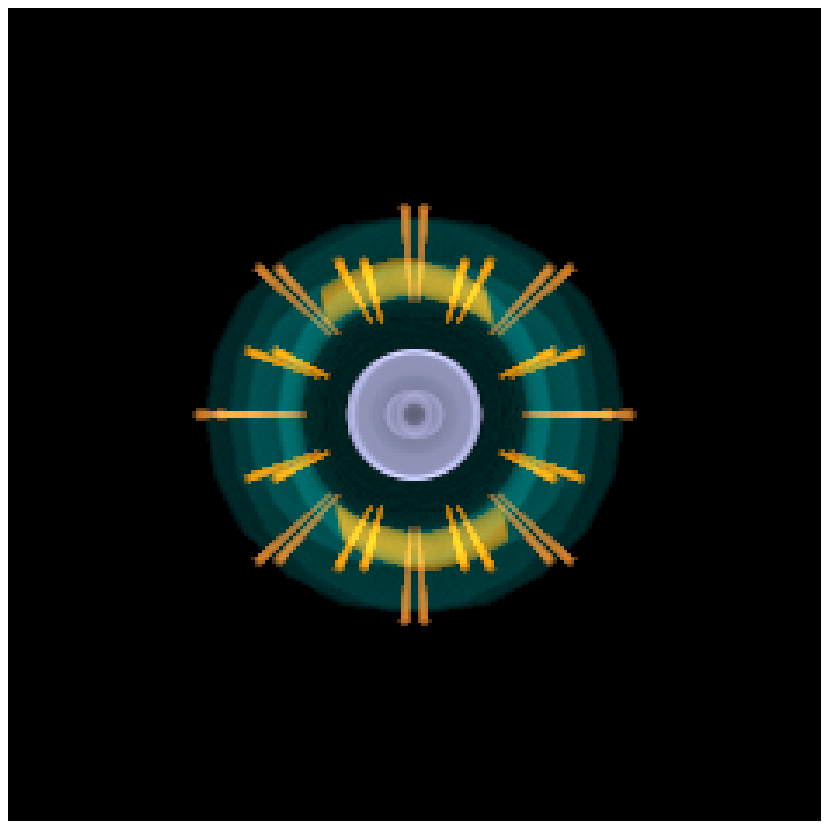}} &
\resizebox{70mm}{!}{\includegraphics{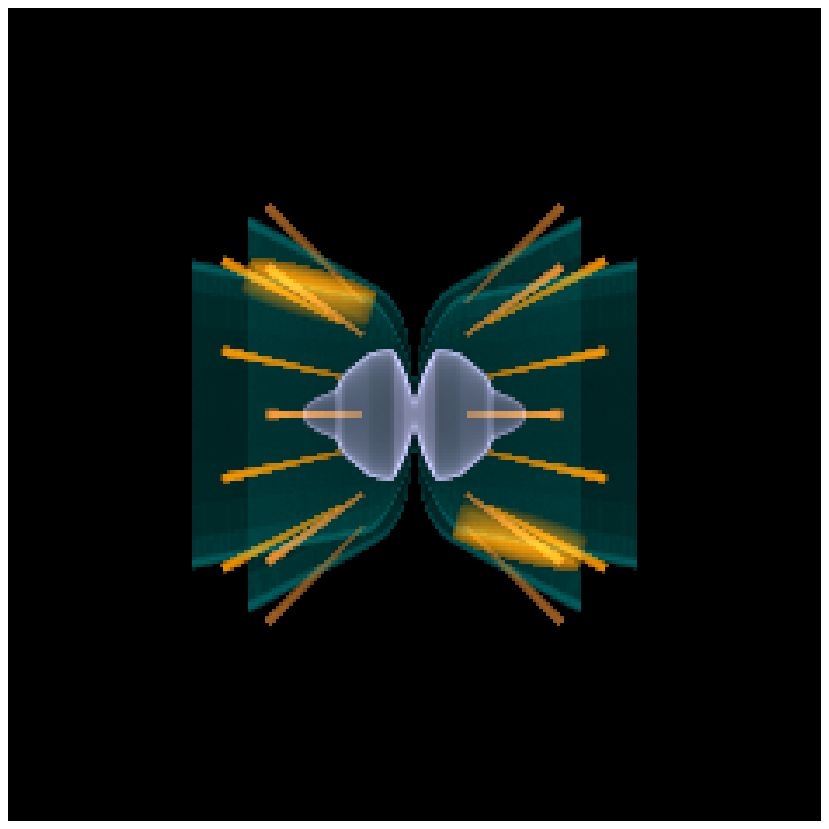}} \\
 \end{tabular}
\caption{SHAPE renderings of the three dimensional model viewed nearly pole-on (left panel) and edge-on (right panel), corresponding to NGC\,2392 and Mz\,3, respectively.
}
  \label{shap}
 \end{center}
 \end{figure}

To test our idea, we construct a model using the software SHAPE version 4.02 \citep{steffen11}, a morpho-kinematic modelling tool used to reconstruct the three-dimensional structures and observed spectral line profiles of gaseous nebulae. The model of the PN consists of four major components: a pair of central lobes, the inner and outer outflows, many knots with tails, and two fuzzes with each adhering to opposite respective surfaces of the inner outflows, as shown in Figure~\ref{mesh}. The density ratio between the central lobes, outflows, knots and fuzzes is assumed
to be 5:2:20:4.  As can be seen from the renderings (Figure~\ref{shap}), this density distribution does a good job of replicating the observations of NGC\,2392 and Mz\,3.  According to our model, the inner and outer regions of NGC\,2392 correspond to the lobes and outflows 
of Mz\,3, respectively, and the bright loops in the inner region may correspond to the overlapping parts of the sub-lobes in Mz\,3. In order to develop the bipolar structure, there might exist an optically thick central disk/torus perpendicular to the outflow axis. Therefore, the nebula is presumably ionization-bounded in the direction perpendicular to the outflow axis and density-bounded in the other directions. This can explain the complex ionization structures of NGC\,2392; the inner region exhibits both high- and low-ionization states, whereas the outer diffuse disk is in a high-ionization state.

\begin{figure*}
\begin{center}
\epsfig{file=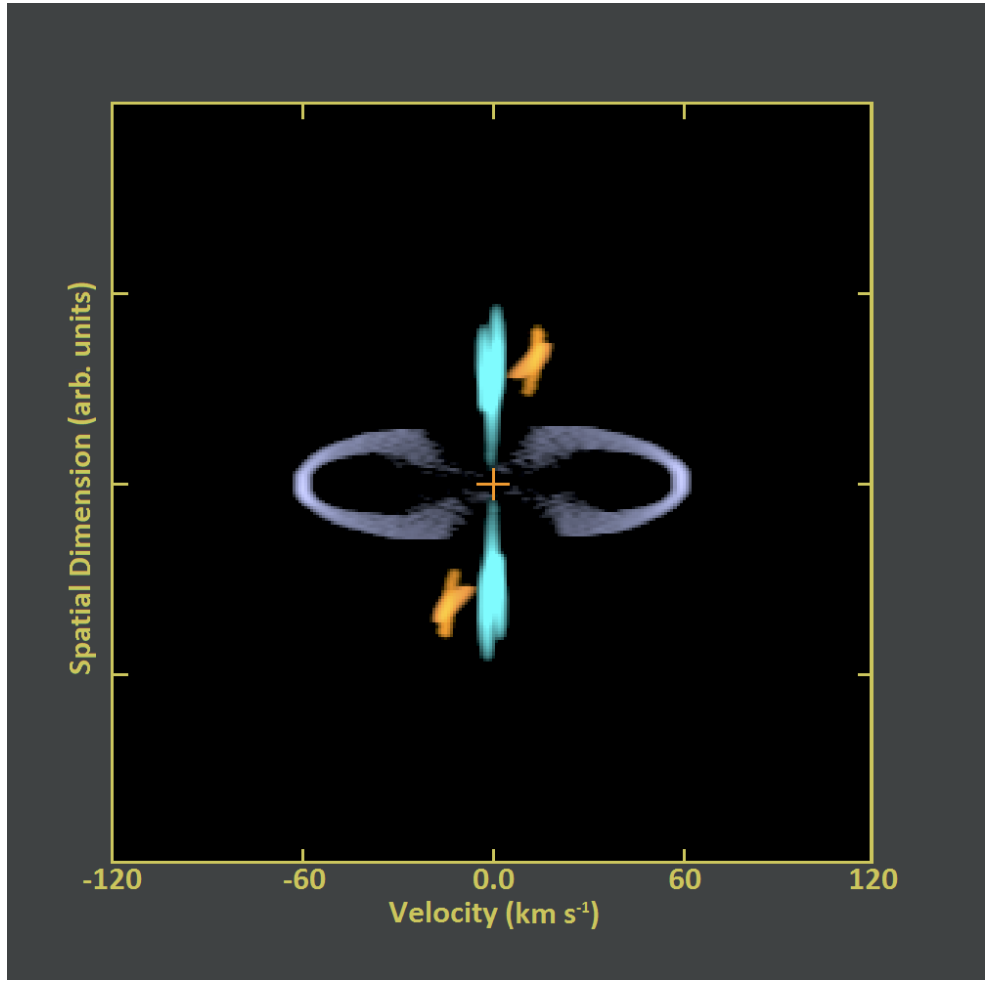,
height=13cm}
\end{center}
\caption{The modelled PV diagram of NGC\,2392 for a slit along the approximate north-south direction and across the central star and the two fuzzes. The colors representing each components are the same as those shown in Figure~\ref{shap}.}
\label{pv}
\end{figure*}

The position-velocity (PV) diagrams of NGC\,2392 can provide important constrains for the modelling. According to the observations of \citet{odell85} and \citet{dufour11}, the main characteristics of the PV diagrams of NGC\,2392 are: 1) the outer region has a low velocity along the line of sight; 2) the inner region has a high velocity along the line of sight; 3) the north and south fuzzes show opposite velocities,  one with a redshift and the other with a blueshift. We simulate the PV diagram of NGC\,2392 for the slit along the approximate north-south direction and across the central star and the two fuzzes. The modelled PV diagram is given in Figure~\ref{pv}, which highly resembles the observations by \citet{odell85} and \citet{dufour11}. For the modelling, the assumed radial velocities of the lobes, outflows, knots, and fuzzes are 150, 55, 90, and 90\,km\,s$^{-1}$, respectively. At a distance of 2.1\,kpc the four components have dynamical ages of 860, 4200, 2600, and 2300 years, respectively, suggesting that the inner region is about five times younger than the outer region.

A variety of models have previously been proposed to explain the appearance of NGC\,2392 (See Section~1).  In the current study we show 
that different perspectives of the proposed structure can mimic the appearance of another well observed PN, suggesting that the morphology 
of NGC\,2392 is not unique. One of the most notable differences between the previous models and ours is the explanation of the outer disk.
We claim that it has a bi-cone structure, and its location in the PV diagram can be attributed to a large opening angle and low expansion velocity. In the image of NGC\,2392 (Figure~\ref{hst}), a striking but barely mentioned feature is an extruded arc-shaped structure in
the east of the outer disk.  This structure is also highly-ionized, and looks like a fainter extension of the outer disk. According to our model, it is a projection of one of the cones, and the outer disk is the overlapping part of the projection of the double cones. It is hard to reproduce such a structure using models (e.g. spherical or elliptical envelope) other than the bi-cone model.  This further supports the validity of our model. \citet{odell02} remarked that the nature of some lowly-ionized features extending outside the outer disk is difficult to explain.  The current model implies that these features are the counterparts of other lowly-ionized features in the faint outskirts of the cones.

 It is well established that the morphology of PNe is statistically related to chemical composition \citep[e.g.][]{sta06,maciel11}. The bipolar nature of NGC\,2392 and Mz\,3 is a general characteristic of Type-I PNe, i.e. PNe with high He/H and N/O abundance ratios \citep{pei83}. Although \citet{per91} classified NGC\,2392 as a Type-I PN, our calculations suggest a low N/O, making it unlikely to be a Type-I PN. A possible cause of this discrepancy is the large error of N abundance, as noted in Section~3.5.1. N$^+$ is the only nitrogen ionic species detected in our optical spectra. If assuming that the N$^+$/N ratio is the same as that derived by \citet{pot08}, we would obtain a significantly larger N/O ratio than that listed in Table.~\ref{element}. On the other hand, the relation between PN abundances and morphology is only statistical and some bipolar PNe indeed exhibit low N/O abundance ratios \citep{pot00}. Therefore, one cannot make a simple statement whether the elemental abundances of NGC\,2392 support or oppose the bipolar hypothesis.

\subsection{Infrared Emission}

\begin{figure*}
\begin{center}
\epsfig{file=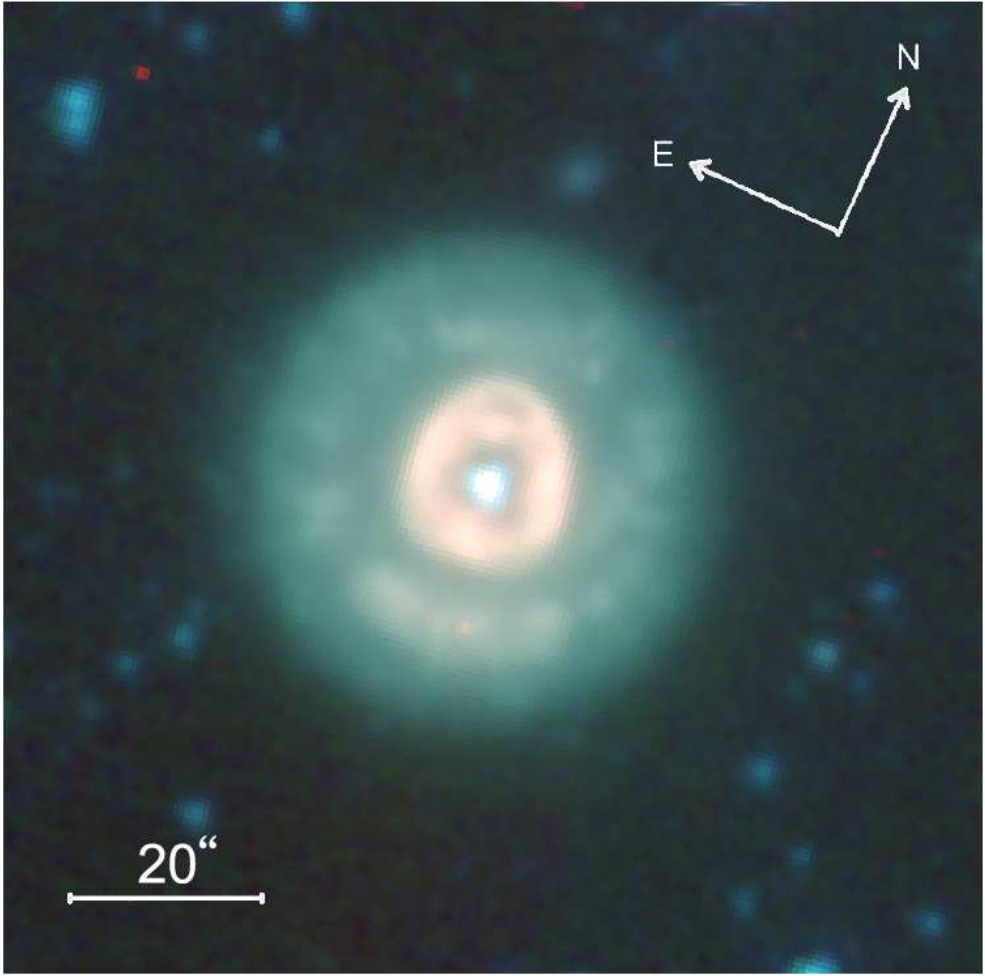,
height=13cm}
\end{center}
\caption{Composite-color images of NGC\,2392 made from three IRAC bands. The 3.6\,$\mu$m is shown as blue, the 4.5\,$\mu$m is shown as green, and the 8.0\,$\mu$m is shown as red.}
\label{irac}
\end{figure*}

In order to further explore the relationship of iron abundance and dust processes, we use the images acquired by the Infrared Array Camera
(IRAC; Fazio et al. 2004) on the {\it Spitzer Space Telescope} as a supplement to our research. The images, centered at approximately 3.6, 4.5, 5.8, and 8.0\,$\mu$m, were obtained from the {\it Spitzer} archive as part of the program No. 30285 (PI: Giovanni Fazio).  In Figure~\ref{irac} we show the composite-color images. The IRAC colors in the two regions are very different. Compared to the 3.6 and 4.5\,$\mu$m bands, the 8.0\,$\mu$m emission is more clumpy in the outer region, and brighter in the inner region. 

\begin{figure*}
\begin{center}
\epsfig{file=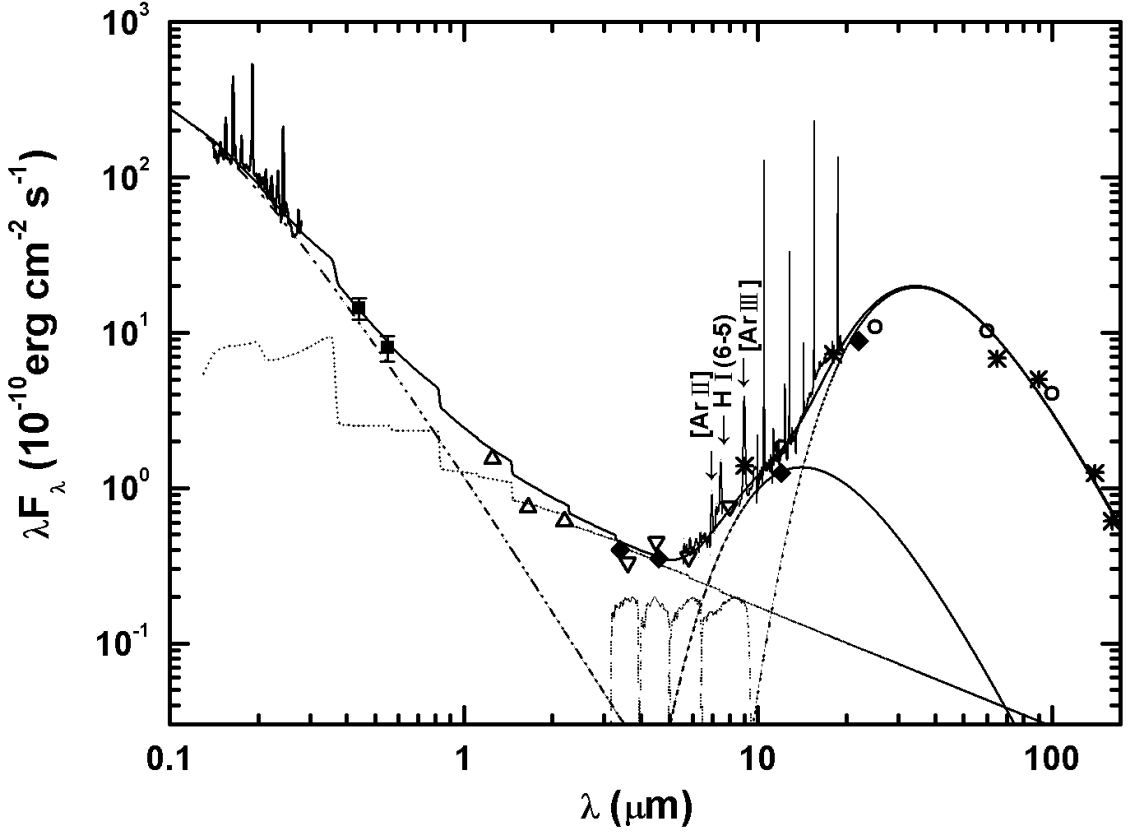,
height=13cm}
\end{center}
\caption{The SED of NGC\,2392 together with an empirical fitting. The filled squares with error bars are the $B$ and $V$ photometry of the central star. The open triangles, open converse triangles, filled lozenges, open circles, and asterisks represent the photometry from 2MASS, 
{\it Spitzer}/IRAC, WISE, {\it IRAS}, and AKARI surveys, respectively. The overlaid spectra (from left to right) were obtained by IUE and 
{\it Spitzer}/IRS. The model curve with step jumps represents the nebular free-bound emission. The three blackbody-like curves (from left to right) represent the central star, and the warm and cool dust components. The total fluxes of all the components are plotted as a solid line on top. The normalized relative spectral response curves for the four {\it Spitzer}/IRAC bands are plotted as dashed lines on the bottom. The three atomic lines which may contribute to the IRAC 8.0\,$\mu$m band are marked.
}
\label{sed}
\end{figure*}

To establish the relative contributions of the different components to these infrared emission bands, we construct a spectral energy distribution (SED) by collecting the photometries from a variety of datasets (see  Zhang et al. 2011), as displayed in Figure~\ref{sed}. The photometries of IRAC and WISE are performed using the same method described in \citet{zhang11}, which yields $F_{3.6, 4.5, 5.8, 8.0\mu{\rm m}}=39$, 72, 69, 202\,mJy and $F_{3.4, 4.6, 12, 22\mu{\rm m}}=45$, 53, 460, 6468\,mJy, respectively. As shown in Figure~\ref{sed}, the overlapping regions between different photometries match well with each other.

The SED clearly shows contributions from different components including photospheric emission, free-bound emission from the gaseous nebula, and dust thermal  emission. The infrared emission of NGC\,2392 can be fit by two modified blackbody curves. The warm and cool dust components have color temperatures of 206 and 84\,K, respectively. According to our fitting, the fraction of total flux between 0.1\,$\mu$m to 170\,$\mu$m
from the photospheric, nebular gaseous continuum emission, and dust components is 71$\%$, 10\,$\%$, and 19$\%$, respectively, 

Figure~\ref{sed} suggests that the IRAC 3.6 and 4.5\,$\mu$m bands are dominated by the nebular free-bound emission. This can be confirmed by the fact that the 3.6 and 4.5\,$\mu$m  images closely resemble the {\it HST} H$\alpha$ image.  Figure~\ref{sed} also indicates that the 8.0\,$\mu$m band mainly traces the warm dust component albeit with minor contributions from atomic lines and nebular continuum. Therefore, in the outer region the close association between the 8.0\,$\mu$m emission and the spatial distribution of lowly-ionized condensations (Figure~\ref{irac}), may be attributed to the fact that the cold and neutral environments favor the survival of dust grains. 

The infrared spectrum of NGC\,2392 does not exhibit broad emission features that might arise from Fe-containing molecules (e.g. FeO and FeS are candidate carriers for the 21\,$\mu$m and 23.5\,$\mu$m bands, respectively). Moreover, the abundances of other elements like oxygen and sulfer do not show significant difference between the two regions. It follows that the depleted iron is unlikely to exist mainly in the form of compounds, but rather in a pure metallic form as suggested by \citet{kemper02}. Lines of molecular hydrogen are also not detected in the infrared spectrum of NGC\,2392.  This is similarly found in Mz\,3 by \citet{smith03}. This behavior distinguishes them from other bipolar PNe since there is a strong correlation between the H$_2$ emission and bipolar structure in PNe \citep{kw96}.

\subsection{The Possible Origin of Iron Abundance Gradient}

Our SHAPE model has shown that the dynamical age of the inner region is smaller than the outer region by a factor of about five. This strengthens
the possibility that the material in the inner region is recently expelled from the giant companion, and thus less iron atoms are depleted into dust grains compared to those in the outer region that has been developed during the pre-PN stage. Furthermore, the detection of X-ray emission \citep{gue05} might indicate that the inner lobes are filled with high-temperature gas, which greatly hinders the condensation of recently ejected
iron atoms. 

The destruction of dust grains by stellar winds and/or radiation might provide an additional mechanism for the iron enhancement in the inner lobes. The precipitation of iron has been considered as nucleation sites for dust formation \citep{lewis79}. If these grains are destroyed by shocked stellar winds, heating by radiation, or grain-grain collisions, iron can be released into the nebular gas. There are dozens of papers discussing the destruction mechanisms of dust grains in various environments \citep[e.g.][]{bar78,ds79,jn11}. Since NGC\,2392 shows high velocities and evidence for shock interaction, sputtering of the grains in shocks might be a plausible scenario in this PN. The destruction of dust grains modifies their size distribution in the inner region which presumably contains a larger fraction of small particles. Because of a low heat capacity, these small grains can be heated to a higher temperature. This in turn explains the remarkable strengthening of 8\,$\mu$m emission in the inner region, as seen in Figure~\ref{irac}.

The investigation of the abundance pattern of other refractory elements, such as Si and Mg, is fundamental to understanding dust processes that modify the abundance of gas-phase iron. \citet{peq80} found a strong gradient of magnesium abundance in the high-excitation PN NGC\,7027. The Mg abundance in the inner part is higher than the outer part by about one order of magnitude.
However, unlike that in NGC\,2392, the iron abundance in NGC\,7027 is uniform.  \citet{peq80} suggested that the most possible explanation of the different behaviors of Mg and Fe is a selective destruction of metallic magnesium grains by radiation heating during an earlier phase of the nebula expansion. Magnesium lines in NGC\,2392 are unfortunately below our detection limits, and so understanding the different behaviors of refractory elements in PNe remains an intriguing challenge for the future.  In addition, unlike that of NGC2392, the infrared spectrum of NGC\,7027  exhibits rich aromatic infrared bands (AIBs). Since the carriers of AIBs have been commonly regarded as building blocks of carbon dust grains, it merits further investigation whether there is a causal link between the different behaviors of AIBs and iron in the two PNe.

\section{Conclusions}

We present the spectra of the inner and outer regions of the PN NGC\,2392. Several [Fe~{\sc iii}] lines are clearly detected. We find that these [Fe~{\sc iii}] lines arise dominantly from the inner region, indicating an enhancement of iron in this region. The spatial distribution of the [Fe~{\sc iii}] lines is similar to that in the bipolar PN Mz\,3, where the [Fe~{\sc iii}] lines are also from the central region. We speculate that NGC\,2392 may be a twin PN of Mz\,3 although they look very different in optical images.  Through a test model, we conclude that it is indeed possible that NGC\,2392 and Mz\,3 have the same intrinsic structure and their distinct appearances are simply due to different orientations relative to the observers' line of sight. The bipolar structure might result from binary interaction. A possible explanation for the Fe abundance gradient is that the Fe atoms in the outer region are mostly condensed into dust grains, whereas the inner region is formed by recently launched stellar winds from the giant companion where less iron atoms are depleted.

Apart from the Eskimo and the Ant PNe, the Dumbbell and the Ring PNe are also found to have the same intrinsic structure viewed from
different perspectives \citep{kc08}. Since it is possible to use identical intrinsic structures to reproduce the markedly different visible morphologies of two of the best-observed PNe, this raises the issue that the current morphology classifications, which are only based on apparent shapes, do not reliably reflect the natural grouping of PNe. Multiwavelength imaging and spectral observations are needed to achieve a more robust PN classification.

Finally, we would like to mention that the behavior of [Fe~{\sc iii}] lines in NGC\,2392 probably suggests a previously unrevealed phenomena in PN evolution. The inner region showing high iron abundance also exhibits strong 8.0\,$\mu$m emission, suggesting that the variations of iron abundance might be related to dust processes. Thus far, iron lines in PNe have been given little attention because they are relatively weak in the majority of PNe. Future observations of emission lines arising from iron and other refractory elements will therefore prove invaluable.  

\acknowledgments

We thank the anonymous referee for critical comments that helped us to significantly improve the paper. We also thank the staff of Gao Meigu Observatory for their help with the observations. Financial support for this work was provided by the Research Grants Council of the Hong Kong under grants HKU7073/11P, the Seed Funding Programme for Basic Research in HKU (200909159007), and the 
Key National Natural Science Foundation of China (No. 10933001).

\clearpage

\clearpage

\begin{deluxetable}{cc@{\extracolsep{0.1in}}ccccccccc}
\tabletypesize{\scriptsize}
\tablecaption{Line fluxes of the YFOSC long-slit spectra on a scale where H$\beta=100$.  \label{lline}}
\tablewidth{0pt}
\tablehead{\multicolumn{2}{c}{$\lambda_{\rm obs}^a$} & \multicolumn{2}{c}{$I^a$} &
{Ion} & {$\lambda_{\rm lab}$} &
{Mult.} &{Lower Term}&{Upper Term}& {$g_1$}&{$g_2$} \\
\cline{1-2}  \cline{3-4}
Outer & Inner & Outer & Inner}
\startdata
3424.36& ...  &  0.69 & ...  &  [Ne~{\sc v}] &  3425.87&         &   2p$^2$ $^3$P  &  2p$^2$ $^1$D &  5&  5\\
3443.80& ...  &  3.10 & ...  &  O~{\sc iii}   &  3444.07& V15    &   3p  $^3$P  &  3d  3P$^{\rm o}$&  5&  5\\
3586.37& ...  &  0.90 & ...  &  He~{\sc i}    &  3587.28& V32    &   2p  $^3$P$^{\rm o}$ &  9d  $^3$D &  9& 15\\
3726.26& 3725.55&179.04 &157.74&  [O~{\sc ii}]  &  3726.03& F1     &   2p$^3$ $^4$S$^{\rm o}$ &  2p$^3$ $^2$D$^{\rm o}$&  4&  4\\
  $*$  & $*$ &  $*$ & $*$  &  [O~{\sc ii}]  &  3728.82& F1     &   2p$^3$ $^4$S$^{\rm o}$ &  2p$^3$ $^2$D$^{\rm o}$&  4&  6\\
3749.49&  ... & 3.70 &  ... &  H~12    &  3750.15& H12    &   2p+ $^2$P$^{\rm o}$ & 12d+ $^2$D &  8&  *\\
3769.94&  ... & 4.44 &  ... &  H~11    &  3770.63& H11    &   2p+ $^2$P$^{\rm o}$ & 11d+ $^2$D &  8&  *\\
3797.35&  ... & 5.65 &  ... &  H~10    &  3797.90& H10    &   2p+ $^2$P$^{\rm o}$ & 10d+ $^2$D &  8&  *\\
3819.90&  ... & 1.14 &  ... &  He~{\sc i}    &  3819.62& V22    &   2p  $^3$P$^{\rm o}$ &  6d  $^3$D &  9& 15\\
3834.98&  ... & 7.41 &  ... &  H~9     &  3835.39& H9     &   2p+ $^2$P$^{\rm o}$ &  9d+ $^2$D &  8&  *\\
3868.53&  3869.01&88.37 & 85.97&  [Ne~{\sc iii}]&  3868.75& F1     &   2p$^4$ $^3$P  &  2p$^4$ $^1$D &  5&  5\\
3888.65&  3889.81&18.72 & 9.62 &  H~8     &  3889.05& H8     &   2p+ $^2$P$^{\rm o}$ &  8d+ $^2$D &  8&  *\\
  $*$  &  $*$ & $*$ & $*$  &   He~{\sc i}    &  3888.65& V2     &   2s  $^3$S  &  3p 3P$^{\rm o}$&  3&  9\\
3923.53&   ... & 0.35 &  ... &  He~{\sc ii}   &  3923.48&  4.15  &   4f+ $^2$F$^{\rm o}$ & 15g+ $^2$G & 32&  *\\
3968.42&  3969.38&34.51 & 21.42&  [Ne~{\sc iii}]&  3967.46& F1     &   2p$^4$ $^3$P  &  2p$^4$ $^1$D &  3&  5\\
  $*$  &  $*$ & $*$ & $*$  &    H~7   &  3970.07& H7     &   2p+ $^2$P$^{\rm o}$ &  7d+ $^2$D &  8& 98\\
4009.21& ... &  0.19 &  ... &  He~{\sc i}    &  4009.26& V55    &   2p  $^1$P$^{\rm o}$ &  7d  $^1$D &  3&  5\\
4026.20& ... &  1.79 &  ... &  He~{\sc i}    &  4026.19& V18    &   2p $^3$P$^{\rm o}$  &  5d  $^3$D & 18&  *\\
4069.15&  4069.96& 2.21 &  2.89&  [S~{\sc ii}]  &  4068.60& F1     &   2p$^3$ $^4$S$^{\rm o}$ &  2p$^3$ 2P$^{\rm o}$&  4&  4\\
4077.25&  4078.19& 0.70 &  1.96&  [S~{\sc ii}]  &  4076.35& F1     &   2p$^3$ $^4$S$^{\rm o}$ &  2p$^3$ 2P$^{\rm o}$&  4&  2\\
4101.92&  4102.97& 22.48 & 15.53&  H~6     &  4101.74& H6     &   2p+ $^2$P$^{\rm o}$ &  6d+ $^2$D &  8& 72\\
4120.57& ...   &  0.17 &...   &  He~{\sc i}    &  4120.84& V16    &   2p  $^3$P$^{\rm o}$ &  5s  $^3$S &  9&  3\\
4144.01& ...   &  0.23 &...   &  He~{\sc i}    &  4143.76& V53    &   2p  $^1$P$^{\rm o}$ &  6d  $^1$D &  3&  5\\
4200.07& ...   &  0.39 &...   &  He~{\sc ii}   &  4199.83&  4.11  &   4f+ $^2$F$^{\rm o}$ & 11g+ $^2$G & 32&  *\\
4268.21& ...   &  0.09 &...   &  C~{\sc ii}    &  4267.00& V6     &   3d  $^2$D  &  4f  $^2$F$^{\rm o}$&  4& 6\\
   $*$ & ...  & $*$   & ...  &  C~{\sc ii}    &  4267.26& V6     &   3d  $^2$D  &  4f  $^2$F$^{\rm o}$&  12& 14\\
4340.86& 4342.17& 48.38 & 39.35&  H~5     &  4340.47& H5     &   2p+ $^2$P$^{\rm o}$ &  5d+ $^2$D &  8& 50\\
4363.66& 4364.58& 15.05 & 14.90&  [O~{\sc iii}] &  4363.23& 2F    &   2p$^2$ $^1$D  &  2p$^2$ $^1$S &  5&  1\\
4388.24&  ...   & 0.49 &...   &  He~{\sc i}    &  4387.93& V51    &   2p  $^1$P$^{\rm o}$ &  5d  $^1$D &  3&  5\\
4471.95&  4472.97& 4.14 &  2.58&  He~{\sc i}    &  4471.50& V14    &   2p  $^3$P$^{\rm o}$ &  4d  $^3$D &  9& 15\\
4542.10&  ... &  0.92 &...   &  He~{\sc ii}   &  4541.59& 4.9    &   4f+ $^2$F$^{\rm o}$ &  9g+ $^2$G & 32&  *\\
4658.66&  4659.31& 0.17 &  2.86&  [Fe~{\sc iii}]&  4658.10& F3     &   3d$^6$ $^5$D  &  3d$^6$ $^3$F2&  9&  9\\
4686.34&  4687.24& 23.90 & 46.60&  He~{\sc ii}   &  4685.68&  3.4   &   3d+ $^2$D  &  4f+ $^2$F$^{\rm o}$& 18& 32\\
  ...  &  4702.64&... &  0.97&  [Fe~{\sc iii}]&  4701.62& F3     &   3d$^6$ $^5$D  &  3d$^6$ $^3$F &  7&  7\\
4712.33& 4713.98&1.94 &  2.02&  [Ar~{\sc iv}] &  4711.37& F1     &   3p$^3$ $^4$S$^{\rm o}$ &  3p$^3$ $^2$D$^{\rm o}$&  4&  6\\
  $*$  & $*$ & $*$ &  $*$    &   He~{\sc i}    & 4713.20 &   V12   &  2p $^3$P$^{\rm o}$   &  4s $^3$S    &  9 &  3\\
4740.78&  4741.98& 1.20 &  1.98&  [Ar~{\sc iv}] &  4740.17& F1     &   3p$^3$ $^4$S$^{\rm o}$ &  3p$^3$ $^2$D$^{\rm o}$&  4&  4\\
  ...  & 4755.21& ... &  0.70&  [Fe~{\sc iii}]&  4754.83&        &   3d$^6$ $^5$D  &  3d$^6$ $^3$F &  7&  9\\
  ...  & 4769.78& ... &  0.40&  [Fe~{\sc iii}]&  4769.60&        &   3d$^6$ $^5$D  &  3d$^6$ $^3$F &  5&  7\\
4862.01& 4862.63&100.00 &100.00&  H~4     &  4861.33& H4     &   2p+ $^2$P$^{\rm o}$ &  4d+ $^2$D &  8& 32\\
  ...  & 4882.10&... &  1.17&  [Fe~{\sc iii}]&  4881.11& F2     &   3d$^6$ $^5$D  &  3d$^6$ $^3$H &  9&  9\\
4922.96& 4922.36& 1.13 &  1.45&  He~{\sc i}    &  4921.93& V48    &   2p  $^1$P$^{\rm o}$ &  4d  $^1$D &  3&  5\\
4959.77& 4960.32&318.93 &304.95&  [O~{\sc iii}] &  4958.91& F1     &   2p$^2$ $^3$P  &  2p$^2$ $^1$D &  3&  5\\
5007.73& 5008.33&962.90 &903.50&  [O~{\sc iii}] &  5006.84& F1     &   2p$^2$ $^3$P  &  2p$^2$ $^1$D &  5&  5\\
5047.36&  ...   & 0.12 &...   &  He~{\sc i}    &  5047.74& V47    &   2p  $^1$P$^{\rm o}$ &  4s  $^1$S &  3&  1\\
5199.43&  ...   & 1.97 &...   &  [N~{\sc i}]   &  5197.51& V1F    &   2p$^3$ $^4$S$^{\rm o}$ &  2p$^3$ $^2$D$^{\rm o}$&  4&  4\\
  $*$  &    ... & $*$  &  ... &  [N~{\sc i}]   &  5200.26& F1     &   2p$^3$ $^4$S$^{\rm o}$ &  2p$^3$ $^2$D$^{\rm o}$&  4&  6\\
5271.23&  5271.66& 0.09 & 1.32 &  [Fe~{\sc iii}]&  5270.40&        &   3d$^6$ $^5$D  &  3d$^6$ $^3$P &  7&  5\\
5411.46&  5410.39& 1.75 &  1.28&  He~{\sc ii}   &  5411.52& 4.7    &   4f+ $^2$F$^{\rm o}$ &  7g+ $^2$G & 32& 98\\
5517.50&  5517.59& 0.78 &  0.68&  [Cl~{\sc iii}]&  5517.66& F1     &   2p$^3$ $^4$S$^{\rm o}$ &  2p$^3$ $^2$D$^{\rm o}$&  4&  6\\
5537.93&  5537.66& 0.68 &  0.64&  [Cl~{\sc iii}]&  5537.60& F1     &   2p$^3$ $^4$S$^{\rm o}$ &  2p$^3$ $^2$D$^{\rm o}$&  4&  4\\
5754.70&  5755.65& 1.76 &  1.53&  [N~{\sc ii}]  &  5754.60& F3     &   2p$^2$ $^1$D  &  2p$^2$ $^1$S &  5&  1\\
5876.18&  5877.24& 8.81 &  8.60&  He~{\sc i}    &  5875.66& V11    &   2p  $^3$P$^{\rm o}$ &  3d  $^3$D &  9& 15\\
6300.87&  6301.27& 3.81 &  0.57&  [O~{\sc i}]   &  6300.34& F1     &   2p$^4$ $^3$P  &  2p$^4$ $^1$D &  5&  5\\
6312.80&  6313.91& 2.41 &  2.48&  [S~{\sc iii}] &  6312.10& F3     &   2p$^2$ $^1$D  &  2p$^2$ $^1$S &  5&  1\\
6364.52&  6364.98& 1.44 &  0.22&  [O~{\sc i}]   &  6363.78& F1     &   2p$^4$ $^3$P  &  2p$^4$ $^1$D &  3&  5\\
6548.46&  6549.06&34.22 & 22.72&  [N~{\sc ii}]  &  6548.10& F1     &   2p$^2$ $^3$P  &  2p$^2$ $^1$D &  3&  5\\
6563.56& 6563.80&224.82 &232.79&  H~3     &  6562.77& H3     &   2p+ $^2$P$^{\rm o}$ &  3d+ $^2$D &  8& 18\\
6583.83& 6584.43&105.72 & 69.46&  [N~{\sc ii}]  &  6583.50& F1     &   2p$^2$ $^3$P  &  2p$^2$ $^1$D &  5&  5\\
6678.97& 6679.22&  2.61 &  2.51&  He~{\sc i}    &  6678.16& V46    &   2p  $^1$P$^{\rm o}$ &  3d  $^1$D &  3&  5\\
6716.74& 6717.62& 13.49 &  4.78&  [S~{\sc ii}]  &  6716.44& F2     &   2p$^3$ $^4$S$^{\rm o}$ &  2p$^3$ $^2$D$^{\rm o}$&  4&  6\\
6731.15& 6732.05& 14.58 &  7.47&  [S~{\sc ii}]  &  6730.82& F2     &   2p$^3$ $^4$S$^{\rm o}$ &  2p$^3$ $^2$D$^{\rm o}$&  4&  4\\
7066.01& 7066.68& 2.13 &  2.54&  He~{\sc i}    &  7065.25& V10    &   2p  $^3$P$^{\rm o}$ &  3s  $^3$S &  9&  3\\
7136.84& 7137.30& 11.73 & 12.44&  [Ar~{\sc iii}]&  7135.80& F1     &   3p$^4$ $^3$P  &  3p$^4$ $^1$D &  5&  5\\
7320.16& 7321.58& 2.93 &  4.13&  [O~{\sc ii}]  &  7318.92& F2     &   2p$^3$ $^2$D$^{\rm o}$ &  2p$^3$ 2P$^{\rm o}$&  6&  2\\
 $*$   &  $*$ & $*$ &  $*$ &  [O~{\sc ii}]  &  7319.99& F2     &   2p$^3$ $^2$D$^{\rm o}$ &  2p$^3$ 2P$^{\rm o}$&  6&  4\\
7330.81&  7331.60&  2.11 &  3.13&  [O~{\sc ii}]  &  7329.67& F2     &   2p$^3$ $^2$D$^{\rm o}$ &  2p$^3$ 2P$^{\rm o}$&  4&  2\\
 $*$   & $*$  & $*$  &  $*$ &  [O~{\sc ii}]  &  7330.73& F2     &   2p$^3$ $^2$D$^{\rm o}$ &  2p$^3$ 2P$^{\rm o}$&  4&  4\\
7752.30& 7752.83& 2.80 &  3.28&  [Ar~{\sc iii}]&  7751.06&        &   3p$^4$ $^3$P  &  3p$^4$ $^1$D &  3&  5\\
8237.92& 8238.07&  0.51 &  0.59&  He~{\sc ii}   &  8236.77& 5.9    &   5g+ $^2$G  &  9h+ $^2$H$^{\rm o}$& 50&  *\\
8578.74& ... &  0.37 &  ... &  [Cl~{\sc ii}] &  8578.70&        &   3p$^4$ $^3$P  &  3p$^4$ $^1$D &  5&  5\\
9016.88& 9017.52&  1.07 &  2.16&  H~{\sc i}     &  9015.00& P10    &   3d+ $^2$D  & 10f+ $^2$F$^{\rm o}$& 18&  *\\
9071.26& 9071.74& 18.51 & 24.31&  [S~{\sc iii}]  &  9069.19&        &   3p2 $^3$P  & 3p2 $^1$D  &  3&  5\\
9231.61& 9232.14& 2.15 &  2.55&  H~{\sc i}     &  9229.70 & P9     &   3d+ $^2$D  &  9f+ $^2$F$^{\rm o}$& 18&  *\\
9532.25& 9532.44& 19.96 & 14.91&  [S~{\sc iii}]  &  9531.10&       &    3p2  $^3$P  &3p2 $^1$D &  5 &5 \\ 
\enddata
\begin{description}
\item $^{a}$ ``$*$'' represents blended lines;
``...'' represents non-detections.
\end{description}
\end{deluxetable}

\begin{deluxetable}{cc@{\extracolsep{0.1in}}ccccccccc}
\tabletypesize{\scriptsize}
\tablecaption{Line fluxes of the YFOSC echelle spectra$^a$ on a scale where H$\beta=100$.  \label{eline}}
\tablewidth{0pt}
\tablehead{\multicolumn{2}{c}{$\lambda_{\rm obs}^b$} & \multicolumn{2}{c}{$I^b$} & 
{Ion} & {$\lambda_{\rm lab}$} &
{Mult.} &{Lower Term}&{Upper Term}& {$g_1$}&{$g_2$} \\
\cline{1-2}  \cline{3-4}
Outer & Inner & Outer & Inner}
\startdata
4685.55& 4685.22 &6.78& 25.62   &  He~{\sc ii}   &  4685.68&  3.4   &   3d+ $^2$D  &  4f+ $^2$F$^{\rm o}$& 18& 32\\
4711.39 & 4711.79&1.16&  2.97   &  [Ar~{\sc iv}] &  4711.37& F1     &   3p$^3$ $^4$S$^{\rm o}$ &  3p$^3$ $^2$D$^{\rm o}$&  4&  6\\
$*$ &  $*$    & $*$ &  $*$    &   He~{\sc i}    & 4713.20 &   V12   &  2p $^3$P$^{\rm o}$   &  4s $^3$S    &  9 &  3\\ 
4741.72&4741.81&0.89&  1.89    & [Ar~{\sc iv}] &  4740.17& F1     &   3p$^3$ $^4$S$^{\rm o}$ &  3p$^3$ $^2$D$^{\rm o}$&  4&  4\\  
4862.92&4862.89&100.00&100.00  &  H~4     &  4861.33& H4     &   2p+ $^2$P$^{\rm o}$ &  4d+ $^2$D &  8& 32\\
4922.79&4922.75& 1.07&  0.76   &  He~{\sc i}    &  4921.93& V48    &   2p  $^1$P$^{\rm o}$ &  4d  $^1$D &  3&  5\\
5015.81&5016.00& 1.32&   0.90  & He~{\sc i}    &  5015.68& V4     &   2s  $^1$S  &  3p  $^1$P$^{\rm o}$&  1&  3\\
  ... & 4658.07&... &  2.72   &  [Fe~{\sc iii}]&  4658.10& F3     &   3d$^6$ $^5$D  &  3d$^6$ $^3$F &  9&  9\\
  ... & 4701.80&... &  0.96   &  [Fe~{\sc iii}]&  4701.62& F3     &   3d$^6$ $^5$D  &  3d$^6$ $^3$F &  7&  7\\
  ... & 4734.14&... &  0.21   &  [Fe~{\sc iii}]&  4733.87&        &   3d$^6$ $^5$D  &  3d$^6$ $^3$F &  5&  5\\
  ... & 4755.60&... &  0.67   &  [Fe~{\sc iii}]&  4754.83&        &   3d$^6$ $^5$D  &  3d$^6$ $^3$F &  7&  9\\
  ... & 4770.25&... &  0.38   &  [Fe~{\sc iii}]&  4769.60&        &   3d$^6$ $^5$D  &  3d$^6$ $^3$F &  5&  7\\
  ... & 4779.81:&... &  0.14:&  [Fe~{\sc iii}]&  4777.88&        &   3d$^6$ $^5$D  &  3d$^6$ $^3$F &  3&  5\\
  ... & 4881.48&... &  1.10   &  [Fe~{\sc iii}]&  4881.11& F2     &   3d$^6$ $^5$D  &  3d$^6$ $^3$H &  9&  9\\
\enddata
\begin{description}
\item $^{a}$ Only the order covering [Fe~{\sc iii}] lines is used. The [Fe~{\sc iii}] lines
are not detected in the outer region.
Note that the [O~{\sc iii}] $\lambda\lambda4959,5007$ lines
are saturated, and are not listed here.
\item $^{b}$ ``$*$'' represents blended lines;
``...'' represents non-detections.
\end{description}
\end{deluxetable}

\begin{deluxetable}{cccccccrr}
\tabletypesize{\scriptsize}
\tablecaption{\label{esolinelist} Line fluxes of the ESO spectrum.
All line fluxes are dereddened and normalized such that H$\beta$
= 100.}
\tablewidth{0pt}
\tablehead{  $\lambda_{\rm obs}^a$ & {$I^a$} &  {Ion} &
{$\lambda_{\rm lab}$} & {Mult.} & {Lower Term} & {Upper Term} & {$g_{1}$}
& {$g_{2}$} }
\startdata
 4010.17 &  0.240  &      He~{\sc i}    & 4009.26 &   V55   &  2p $^1$P$^{\rm o}$   &  7d $^1$D    &  3 &  5\\
 4027.27 &  1.937  &      He~{\sc i}    & 4026.20 &   V18   &  2p $^3$P$^{\rm o}$   &  5d $^3$D    &  9 & 15\\
 4069.93 &  2.443  &     $[$S~{\sc ii}$]$   & 4068.60 &   F1    &  3p$^3$ $^4$S$^{\rm o}$  &  3p$^3$ $^2$P$^{\rm o}$  &  4 &  4\\
 4077.61 &  0.780  & $[$S~{\sc ii}$]$   & 4076.35 &   F1    &  3p$^3$ $^4$S$^{\rm o}$  &  3p$^3$ $^2$P$^{\rm o}$  &  4 &  2\\
 4098.85 &  0.288  &      N~{\sc iii}   & 4097.33 &   V1    &  3s $^2$S    &  3p $^2$P$^{\rm o}$   &  2 &  4\\
 4102.81 &  0.259  &      H~{\sc i}     & 4101.73 &   H6    &  2p+ $^2$P$^{\rm o}$  &  6d+ $^2$D   &  8 & 72\\
 4121.89 &  0.235  &      He~{\sc i}    & 4120.99 &   V16   &  2p $^3$P$^{\rm o}$   &  5s $^3$S    &  9 &  3\\
 4144.80 &  0.431  &      He~{\sc i}    & 4143.76 &   V53   &  2p $^1$P$^{\rm o}$   &  6d $^1$D    &  3 &  5\\
 4182.12 &  0.096  &    $[$Fe~{\sc v}$]$   & 4180.60 &         &  3d$^4$ $^5$D   &  3d$^4$ $^3$P2  &  3 &  1\\
 4200.96 &  0.533  &     He~{\sc ii}   & 4199.83 &   4.11  &  4f+ $^2$F$^{\rm o}$  & 11g+ $^2$G   & 32 &  *\\
 4228.63 &  0.765  &    $[$Fe~{\sc v}$]$   & 4227.19 &         &  3d$^4$ $^5$D   &  3d$^4$ $^3$H   &  9 &  9\\
 4245.46 &  0.069  &    $[$Fe~{\sc ii}$]$  & 4243.97 &         &  3d$^7$ $^4$F   &  4s $^4$G    & 10 & 12\\
 4268.33 &  0.078  &      C~{\sc ii}    & 4267.26 &   V6    &  3d $^2$D    &  4f $^2$F$^{\rm o}$   &  6 &  8\\
 4341.65 & 46.972  &      H~{\sc i}     & 4340.46 &   H5    &  2p+ $^2$P$^{\rm o}$  &  5d+ $^2$D   &  8 & 50\\
 4364.38 & 19.355  &     $[$O~{\sc iii}$]$  & 4363.21 &   F2    &  2p$^2$ $^1$D   &  2p$^2$ $^1$S   &  5 &  1\\
 4389.16 &  0.406  &      He~{\sc i}    & 4387.93 &   V51   &  2p $^1$P$^{\rm o}$   &  5d $^1$D    &  3 &  5\\
 4394.55 &  0.368  &      [Fe~{\sc ii}$]$ & 4393.07 &         &  3d$^7$ $^6$D   &  3d$^7$ z$^4$Go &  8 &  8\\
 4438.95 &  0.043  &      He~{\sc i}    & 4437.55 &   V50   &  2p $^1$P$^{\rm o}$   &  5s $^1$S    &  3 &  1\\
 4472.75 &  3.479  &      He~{\sc i}    & 4471.50 &   V14   &  2p $^3$P$^{\rm o}$   &  4d $^3$D    &  9 & 15\\
 4542.84 &  1.004  &      He~{\sc ii}   & 4541.59 &   4.9   &  4f+ $^2$F$^{\rm o}$  &  9g+ $^2$G   & 32 &  *\\
 4564.04 &  0.079  &      Mg~{\sc i}$]$   & 4562.60 &         &  3s2 $^1$S   &  3s3p $^3$P$^{\rm o}$ &  1 &  5\\
 4572.77 &  0.221  &      Mg~{\sc i}$]$   & 4571.10 &         &  3s2 $^1$S   &  3s3p $^3$P$^{\rm o}$ &  1 &  3\\
 4608.61 &  0.152  &     $[$Fe~{\sc iii}$]$ & 4607.03 &         &  3d$^6$ $^5$D   &  3d$^6$ $^3$F   &  9 &  7\\
 4635.44 &  0.180  &      N~{\sc iii}   & 4634.14 &   V2    &  3p $^2$P$^{\rm o}$   &  3d $^2$D    &  2 &  4\\
 4642.05 &  0.494  &      N~{\sc iii}   & 4640.64 &   V2    &  3p $^2$P$^{\rm o}$   &  3d $^2$D    &  4 &  6\\
   $*$   &     $*$ &     O~{\sc ii}    & 4641.81 &   V1    &  3s $^4$P    &  3p $^4$D$^{\rm o}$   &  4 &  6\\
   $*$   &     $*$ &     N~{\sc iii}   & 4641.85 &   V2    &  3p $^2$P$^{\rm o}$   &  3d $^2$D    &  4 &  4\\
 4650.77 &  0.131  &     O~{\sc ii}    & 4649.13 &   V1    &  3s $^4$P    &  3p $^4$D$^{\rm o}$   &  6 &  8\\
   $*$    &     $*$ &    C~{\sc iii}   & 4650.25 &    V1     &   3s  3S  &  3p  3P*&  3&  3\\
   $*$    &     $*$ &     O~{\sc ii}    & 4650.84 &   V1    &  3s $^4$P    &  3p $^4$D$^{\rm o}$   &  2 &  2\\
  4659.61 & 1.957  &     $[$Fe~{\sc iii}$]$ & 4658.05 &   F3    &  3d$^6$ $^5$D   &  3d$^6$ $^3$F  &  9 &  9\\
  4668.86 & 0.058  &   $[$Fe~{\sc iii}$]$ & 4667.01 &   F3    &  3d$^6$ $^5$D   &  3d$^6$ $^3$F  &  7 &  5\\
  4687.03 &29.745  &    He~{\sc ii}   & 4685.68 &   3.4   &  3d+ $^2$D   &  4f+ $^2$F$^{\rm o}$  & 18 & 32\\
  4703.13 & 0.590  &   $[$Fe~{\sc iii}$]$ & 4701.53 &   F3    &  3d$^6$ $^5$D   &  3d$^6$ $^3$F  &  7 &  7\\
  4712.82 & 2.865  &   $[$Ar IV$]$  & 4711.37 &   F1    &  3p$^3$ $^4$S$^{\rm o}$  &  3p$^3$ $^2$D$^{\rm o}$  &  4 &  6\\
   $*$    &    $*$ &     He~{\sc i}    & 4713.20 &   V12   &  2p $^3$P$^{\rm o}$   &  4s $^3$S    &  9 &  3\\
  4726.21 & 0.287  &     $[$Ne IV$]$  & 4724.17 &   F1    &  2p$^3$ $^2$D$^{\rm o}$  &  2p$^3$ $^2$P$^{\rm o}$  &  4 &  4\\
   $*$    &    $*$ &    $[$Ne IV$]$  & 4725.67 &   F1    &  2p$^3$ $^2$D$^{\rm o}$  &  2p$^3$ $^2$P$^{\rm o}$  &  4 &  2\\
  4735.56 & 0.225  &    $[$Fe~{\sc iii}$]$ & 4733.91 &   F3    &  3d$^6$ $^5$D   &  3d$^6$ $^3$F  &  5 &  5\\
  4741.60 & 1.898  &    $[$Ar IV$]$  & 4740.17 &   F1    &  3p$^3$ $^4$S$^{\rm o}$  &  3p$^3$ $^2$D$^{\rm o}$  &  4 &  4\\
  4756.35 & 0.360  &    $[$Fe~{\sc iii}$]$ & 4754.69 &   F3    &  3d$^6$ $^5$D   &  3d$^6$ $^3$F  &  7 &  9\\
  4771.05 & 0.251  &    $[$Fe~{\sc iii}$]$ & 4769.43 &   F3    &  3d$^6$ $^5$D   &  3d$^6$ $^3$F  &  5 &  7\\
  4779.34 & 0.101  &    $[$Fe~{\sc iii}$]$ & 4777.68 &   F3    &  3d$^6$ $^5$D   &  3d$^6$ $^3$F  &  3 &  5\\
  4862.83 &100.000  &     H~{\sc i}     & 4861.32 &   H4    &  2p+ $^2$P$^{\rm o}$  &  4d+ $^2$D   &  8 & 32\\
  4882.80 & 0.825  &    $[$Fe~{\sc iii}$]$ & 4881.00 &   F2    &  3d$^6$ $^5$D   &  3d$^6$ $^3$H   &  9 &  9\\
  4904.18 & 0.100  &    $[$Fe~{\sc iv}$]$  & 4903.07 &         &  3d$^5$ $^4$G   &  3d$^5$ $^4$F   &  8 &  8\\
  4907.96 & 0.112  &    $[$Fe~{\sc iv}$]$  & 4906.56 &         &  3d$^5$ $^4$G   &  3d$^5$ $^4$F   & 12 & 10\\
  4923.56 & 0.962  &     He~{\sc i}    & 4921.93 &   V48   &  2p $^1$P$^{\rm o}$   &  4d $^1$D    &  3 &  5\\
  4932.79 & 0.265  &    $[$O~{\sc iii}$]$  & 4931.23 &   F1    &  2p$^2$ $^3$P   &  2p$^2$ $^1$D   &  1 &  5\\
  4960.40 & 383.333 &  $[$O~{\sc iii}$]$  & 4958.91 &   F1    &  2p$^2$ $^3$P   &  2p$^2$ $^1$D   &  3 &  5\\
\enddata
\begin{description}
\item $^a$ ``$*$'' represents blended lines.
\end{description}
\end{deluxetable}

\begin{deluxetable}{lc@{\extracolsep{0.1in}}cc@{\extracolsep{0.1in}}cc}
\tablecaption{Comparison of observed and predicted intensity ratios of pairs of
[Fe~{\sc iii}] lines coming from the same upper level.\label{comp}}
\tablewidth{0pt}
\tablehead{
\multicolumn{2}{c}{Ratio} & \multicolumn{2}{c}{Observed}  & \multicolumn{2}{c}{Predicted}\\
\cline{1-2}  \cline{3-4} \cline{5-6}
Transition  &Wavelength&   YFOSC    &   ESO      &  Nahar (1996) & Quinet(1996) }
\startdata
$\frac{a^3F_4-^5D_4}{a^3F_4-^5D_3}$&$\frac{4658}{4754}$&4.060&5.436&5.333&5.495\\
$\frac{a^3F_2-^5D_1}{a^3F_2-^5D_2}$&$\frac{4777}{4733}$&0.667&0.449&0.483&0.482\\
$\frac{a^3F_3-^5D_3}{a^3F_3-^5D_2}$&$\frac{4701}{4769}$&2.526&2.351&2.895&2.936\\
$\frac{a^3F_3-^5D_4}{a^3F_3-^5D_3}$&$\frac{4607}{4701}$&\nodata&0.258&0.184&0.172\\
$\frac{a^3F_2-^5D_3}{a^3F_2-^5D_2}$&$\frac{4667}{4733}$&\nodata&0.258&0.287&0.276\\
\enddata
\end{deluxetable}

\begin{deluxetable}{lcc}
\tablecaption{Plasma diagnostics based on the YFOSC spectra. \label{diagnostic}}
\tablewidth{0pt}
\tablehead{
{Diagnostic}     & {Outer region} & {Inner region}  }
\startdata
  &\multicolumn{2}{c}{$T_{\rm e}$\,(K)}\\
{\rm [O~{\sc iii}]} $(\lambda5007+\lambda4959)/\lambda4363$ & 13700 & 14000 \\
{\rm [O~{\sc ii}]}  $(\lambda7320+\lambda7330)/(\lambda3726+\lambda3729)$&9300 &10000\\
{\rm [N~{\sc ii}]}  $(\lambda6548+\lambda6584)/\lambda5754$&  10400 & 11800\\
BJ/H\,11 &  12000 & ... \\
  &\multicolumn{2}{c}{$\log N_{\rm e}$\,(cm$^{-3}$)}\\
{\rm [S~{\sc ii}]}  $\lambda6731/\lambda6716$ & 2.94 & 3.52\\
{\rm [Cl~{\sc iii}]}$\lambda5537/\lambda5517$ & 3.02 & 3.20\\
{\rm [Fe~{\sc iii}]} $\lambda4658/\lambda4701$& ...  & 3.80 \\
\enddata
\end{deluxetable}

\begin{deluxetable}{lllll}
\tablecaption{Ionic abundances derived from the YFOSC spectra.
\label{abund}}
\tablewidth{0pt}
\tablehead{
Ion  &    Lines  & \multicolumn{2}{c}{X$^{i+}$/H$^+$} &  Note \\
\cline{3-4} 
& &                  Outer    &      Inner & }
\startdata
He$^+$ & adopted  & 6.950$\times$10$^{-2}$ & 6.152$\times$10$^{-2}$ &  \\
       & He~{\sc i} $\lambda$4471  & 8.426$\times$10$^{-2}$ & 5.251$\times$10$^{-2}$ & long-slit\\
       & He~{\sc i} $\lambda$5876  & 6.487$\times$10$^{-2}$ & 6.333$\times$10$^{-2}$ & long-slit\\
       & He~{\sc i} $\lambda$6678  & 6.171$\times$10$^{-2}$ & 6.459$\times$10$^{-2}$ & long-slit\\
He$^{2+}$  & He~{\sc ii} $\lambda$4686  & 1.960$\times$10$^{-2}$ & 3.823$\times$10$^{-2}$ & long-slit\\
C$^{2+}$ & C~{\sc ii} $\lambda$4267 & 8.822$\times$10$^{-5}$ &  \nodata &  long-slit\\
N$^0$  &$[$N~{\sc i}$]$      $\lambda\lambda$5198,5200 & 1.746$\times$10$^{-6}$ &    \nodata & long-slit\\
N$^{+}$ & adopted &     2.175$\times$10$^{-5}$ & 1.219$\times$10$^{-5}$ & \\
  &$[$N~{\sc ii}$]$     $\lambda\lambda$6548,6584 & 2.166$\times$10$^{-5}$ & 1.212$\times$10$^{-5}$ & long-slit\\
  &$[$N~{\sc ii}$]$     $\lambda$5754      & 2.869$\times$10$^{-5}$ & 1.656$\times$10$^{-5}$ & long-slit\\
O$^0$  &$[$O~{\sc i}$]$      $\lambda\lambda$6300,6364 & 1.038$\times$10$^{-5}$ & 1.185$\times$10$^{-6}$ & long-slit\\
O$^+$  & adopted  & 1.944$\times$10$^{-4}$ & 1.350$\times$10$^{-4}$ & \\
  &$[$O~{\sc ii}$]$     $\lambda\lambda$7320,7330 & 1.081$\times$10$^{-4}$ & 6.818$\times$10$^{-5}$ & long-slit\\
  &$[$O~{\sc ii}$]$     $\lambda$3726      & 1.968$\times$10$^{-4}$ & 1.381$\times$10$^{-4}$ & long-slit\\
O$^{2+}$& adopted & 1.320$\times$10$^{-4}$ &  1.175$\times$10$^{-4}$ & \\
  &$[$O~{\sc iii}$]$    $\lambda\lambda$4959,5007 & 1.320$\times$10$^{-4}$ & 1.175$\times$10$^{-4}$ & long-slit\\
  &$[$O~{\sc iii}$]$    $\lambda$4363      & 1.317$\times$10$^{-4}$ & 1.156$\times$10$^{-4}$ & long-slit\\
Ne$^{2+}$  &$[$Ne~{\sc iii}$]$   $\lambda\lambda$3869,3967 & 9.548$\times$10$^{-5}$ & 4.187$\times$10$^{-5}$ & long-slit\\
Ne$^{4+}$  &$[$Ne~{\sc v}$]$     $\lambda$3426      & 2.097$\times$10$^{-7}$ &      \nodata & long-slit\\
Ar$^{2+}$  &$[$Ar~{\sc iii}$]$   $\lambda\lambda$7136,7751 & 5.089$\times$10$^{-7}$ & 5.301$\times$10$^{-7}$ & long-slit\\
Ar$^{3+}$& adopted & 1.233$\times$10$^{-7}$ & 1.534$\times$10$^{-7}$ & \\
  &$[$Ar~{\sc iv}$]$    $\lambda\lambda$4711,4740 & 1.233$\times$10$^{-7}$ & 1.534$\times$10$^{-7}$ & long-slit\\
  &$[$Ar~{\sc iv}$]$    $\lambda\lambda$4711,4740 & 8.052$\times$10$^{-8}$ & 1.864$\times$10$^{-7}$ & echelle\\
S$^+$ & adopted &  8.771$\times$10$^{-7}$ &  5.266$\times$10$^{-7}$ & \\
  &$[$S~{\sc ii}$]$     $\lambda\lambda$4068,4076 & 8.036$\times$10$^{-7}$ & 7.408$\times$10$^{-7}$ & long-slit\\
  &$[$S~{\sc ii}$]$     $\lambda\lambda$6716,6731 & 8.847$\times$10$^{-7}$ & 4.418$\times$10$^{-7}$ & long-slit\\
S$^{2+}$ & adopted & 1.956$\times$10$^{-6}$ & 1.639$\times$10$^{-6}$ & \\
  &$[$S~{\sc iii}$]$    $\lambda\lambda$9069,9531 & 1.602$\times$10$^{-6}$ & 1.393$\times$10$^{-6}$ & long-slit\\
  &$[$S~{\sc iii}$]$    $\lambda$6312      & 7.607$\times$10$^{-6}$ & 5.532$\times$10$^{-6}$ & long-slit\\
Cl$^+$  &$[$Cl~{\sc ii}$]$    $\lambda$8579      & 4.238$\times$10$^{-8}$ &     \nodata & long-slit\\
Cl$^{2+}$  &$[$Cl~{\sc iii}$]$   $\lambda\lambda$5518,5538 & 1.160$\times$10$^{-7}$ & 8.936$\times$10$^{-8}$ & long-slit\\
Fe$^{2+}$  & adopted &   9.145$\times$10$^{-8}$ &  1.369$\times$10$^{-6}$  & \\
&$[$Fe~{\sc iii}$]$   $\lambda$4658  & 9.832$\times$10$^{-8}$ & 1.405$\times$10$^{-6}$ & long-slit\\
           &$[$Fe~{\sc iii}$]$   $\lambda$4702  &       \nodata &  1.362$\times$10$^{-6}$ & long-slit\\
           &$[$Fe~{\sc iii}$]$   $\lambda$4755  &      \nodata & 1.837$\times$10$^{-6}$ & long-slit\\
           &$[$Fe~{\sc iii}$]$   $\lambda$4770  &      \nodata & 1.624$\times$10$^{-6}$ & long-slit\\
           &$[$Fe~{\sc iii}$]$   $\lambda$4881  &        \nodata & 8.509$\times$10$^{-7}$ & long-slit\\
           &$[$Fe~{\sc iii}$]$   $\lambda$5270  &  7.847$\times$10$^{-8}$ & 7.000$\times$10$^{-7}$ & long-slit\\
  &$[$Fe~{\sc iii}$]$   $\lambda$4658      &      \nodata & 1.337$\times$10$^{-6}$ & echelle\\
  &$[$Fe~{\sc iii}$]$   $\lambda$4702      &       \nodata & 1.348$\times$10$^{-6}$ & echelle\\
  &$[$Fe~{\sc iii}$]$   $\lambda$4734      &       \nodata & 1.133$\times$10$^{-6}$ & echelle\\
  &$[$Fe~{\sc iii}$]$   $\lambda$4755      &      \nodata & 1.758$\times$10$^{-6}$ & echelle\\
  &$[$Fe~{\sc iii}$]$   $\lambda$4770      &      \nodata & 1.543$\times$10$^{-6}$ & echelle\\
  &$[$Fe~{\sc iii}$]$   $\lambda$4778      &       \nodata & 7.625$\times$10$^{-7}$ & echelle\\
  &$[$Fe~{\sc iii}$]$   $\lambda$4881      &        \nodata & 8.000$\times$10$^{-7}$ & echelle\\
\enddata
\end{deluxetable}

\begin{deluxetable}{lll}
\tablecaption{\label{eso_abund} Ionic abundances derived from the ESO spectrum.}
\tablewidth{0pt}
\tablehead{{Ion} & {Lines} & {X$^{i+}$/H$^+$}}
\startdata
He$^{+}$  & He~{\sc i} $\lambda$4471  & 7.081$\times$10$^{-2}$\\
He$^{2+}$ & He~{\sc ii} $\lambda$4686 & 2.588$\times$10$^{-2}$\\
C$^{2+}$  & C~{\sc ii} $\lambda$4267  & 6.641$\times$10$^{-5}$\\
O$^{2+}$  & O~{\sc ii} $\lambda$4649  & 2.179$\times$10$^{-4}$\\
O$^{2+}$  & $[$O~{\sc iii}$]$ $\lambda$,$\lambda$4959,\,4363    & 1.323$\times$10$^{-4}$\\
S$^{+}$  & $[$S~{\sc ii}$]$  $\lambda\lambda$4069,\,4076 & 3.945$\times$10$^{-7}$\\
Mg$^0$    &    Mg~{\sc i}$]$  $\lambda\lambda$4563,\,4571 & 5.121$\times$10$^{-9}$\\
Ar$^{3+}$ & $[$Ar~{\sc iv}$]$ $\lambda\lambda$4711,\,4740 & 1.622$\times$10$^{-7}$\\
Ne$^{3+}$ & $[$Ne~{\sc iv}$]$ $\lambda\lambda$4724,\,4726 & 1.232$\times$10$^{-5}$\\
Fe$^{2+}$ &  adopted                                      & 5.282$\times$10$^{-7}$\\ 
          & $[$Fe~{\sc iii}$]$ $\lambda$4607.03           & 7.207$\times$10$^{-7}$\\
          & $[$Fe~{\sc iii}$]$ $\lambda$4658.05           & 5.131$\times$10$^{-7}$\\
          & $[$Fe~{\sc iii}$]$ $\lambda$4667.01           & 6.016$\times$10$^{-7}$\\
          & $[$Fe~{\sc iii}$]$ $\lambda$4701.53           & 5.217$\times$10$^{-7}$\\
          & $[$Fe~{\sc iii}$]$ $\lambda$4733.91           & 6.704$\times$10$^{-7}$\\
          & $[$Fe~{\sc iii}$]$ $\lambda$4754.69           & 5.034$\times$10$^{-7}$\\
          & $[$Fe~{\sc iii}$]$ $\lambda$4769.43           & 6.426$\times$10$^{-7}$\\
          & $[$Fe~{\sc iii}$]$ $\lambda$4777.68           & 6.227$\times$10$^{-7}$\\
          & $[$Fe~{\sc iii}$]$ $\lambda$4881.00           & 4.538$\times$10$^{-7}$\\
\enddata
\end{deluxetable}

\begin{deluxetable}{lllllll}
\tablecaption{Elemental abundances,
in units such as $\log N$(H)$=12.0$.
\label{element}}
\tablewidth{0pt}
\tablehead{
Element  & \multicolumn{2}{c}{Present} & Pottasch$^a$ &  Henry$^b$ & Barker$^c$ & Solar$^d$ \\
\cline{2-3}
 &                Outer    &      Inner & }
\startdata
He & 10.95 &  10.99  & 10.90 & 10.88 &  10.99 & 10.90\\
C  & 8.11  &  \nodata& 8.52  & 8.34  &  7.62  & 8.39 \\
N  & 7.63: &  7.50:  & 8.27  & 8.04  &  8.04  & 7.83 \\
O  & 8.59  &  8.54   & 8.46  & 8.45  &  8.53  & 8.69 \\
Ne & 8.15  &  8.10   & 7.92  & 7.81  &  7.88  & 7.87\\
S  & 6.47  &  6.37   & 6.70  &\nodata&  6.63  & 7.19 \\
Cl & 5.20  &  5.12   & 5.11  &\nodata&\nodata & 5.26 \\
Ar & 6.09  &  6.25   & 6.34  &\nodata&  6.15  & 6.55 \\
Fe & 5.25  &  6.53   & 5.90  &\nodata&\nodata & 7.47 \\
\enddata
\begin{description}
\item $^{a}$ \citet{pot08};
\item $^{b}$ \citet{henry00};
\item $^{c}$ \citet{barker91};
\item $^{d}$ \citet{lodders}.
\end{description}
\end{deluxetable}

\end{document}